# Consideration of the Intricacies Inherent in Molecular Beam Epitaxy of the NaCl/GaAs System

Brelon J. May[1], Jae Jin Kim[2], Patrick Walker[1], William E. McMahon[1], Helio R. Moutinho[1], Aaron J. Ptak[1], David L. Young[1]

[1]National Renewable Energy Laboratory, Golden, USA
[2]Shell International Exploration and Production Inc., Houston, USA

**ABSTRACT**

The high cost of substrates for III-V growth can be cost limiting for technologies that require large semiconductor areas. Thus, being able to separate device layers and reuse the original substrate is highly desirable, but existing techniques to lift off a film from a substrate have substantial drawbacks. This work discusses some of the complexities with the growth of a water-soluble, alkali-halide salt thin film between a III-V substrate and overlayer. Much of the difficulty stems from the growth of GaAs on an actively decomposing NaCl surface at elevated temperatures. Interestingly, the presence of an *in-situ* electron beam incident on the NaCl surface, prior to and during GaAs deposition, affects the crystallinity and morphology of the III-V overlayer. Here we investigate a wide range of growth temperatures and the timing of the impinging flux of both elemental sources and high energy electrons at different points during the growth. We show that an assortment of morphologies (discrete islands, porous material, and fully dense layers with sharp interfaces) and crystallinity (amorphous, crystalline, and highly textured) occur depending on the specific growth conditions, driven largely by changes in GaAs nucleation which is greatly affected by the presence of the reflection high energy electron diffraction beam.

1. ## INTRODUCTION

Substrate recycling is an area of high interest, especially for technologies that employ expensive single-crystalline materials such as epitaxial III-Vs. The cost of substrates is still problematic for technologies requiring large area such as high efficiency solar cells,[1] while processing costs can be reduced through scaling, and growth costs are being reduced with technologies such as high growth rate metal organic chemical vapor deposition (MOCVD)[2,3] and hydride vapor phase epitaxy (HVPE).[4–6] A number of methods have been demonstrated for separation of a III-V device layer from the parent wafer that can be thought of as either spalling (mechanical removal)[7–9] or wet etching of a sacrificial layer (chemical removal).[10,11] Both techniques have advantages and disadvantages pertaining to the material(s) used, the fracture planes, and the substrate orientation used. It is paramount that the surface remains as pristine as possible to obtain the benefit of cost effectiveness and existing methods on commonly used GaAs (001) substrates fall short. For spalling, there is a mismatch in the desired crack propagation direction and the low-energy cleavage planes resulting in facets at a high angle to the surface normal when spalling GaAs (001) substrates. Existing sacrificial layers for wet-etching techniques (Al-rich III-Vs) also tend to leave behind a rougher surface, and insoluble byproducts on the surface.[12] This work explores the integration of NaCl as a different selective etch layer material in an effort to preserve the epi-ready surface of the parent wafer post exfoliation.

The high solubility of NaCl, and low solubility of III-Vs, in water presents an alternative avenue for the possibility of a rapidly dissolvable sacrificial layer with high selectivity in a non-toxic environment, and was shown to be effective in various material systems and devices.[13–15] Both GaAs and NaCl are cubic, and while NaCl has a higher thermal expansion coefficient, they are lattice matched at ~100°C. The first integration of these two materials occurred in the early days of epitaxy and vacuum deposition, with NaCl being a popular substrate choice for both semiconductors and metals.[16–23] Previous demonstrations of growth of GaAs on NaCl used bulk NaCl substrates that suffered from reactivity with water vapor in the air. These substrates were vacuum-cleaved and deliberately desorbed large amounts of material *in-situ* prior to growth to produce a clean surface.[24] However, bulk NaCl substrates are likely not economical for device exfoliation because the substrate would necessarily be dissolved after each growth. A more elegant approach is to deposit an epitaxial NaCl liftoff layer directly on a GaAs substrate to facilitate liftoff and substrate reuse. This requires the development of the growth of both NaCl on GaAs and GaAs on NaCl. This is not a trivial combination because of the possibility of forming anti-phase domains and twin boundaries when growing GaAs on NaCl because of the higher symmetry of NaCl. The most challenging obstacle is that NaCl is highly volatile at typical GaAs growth temperatures approaching 600°C. Thin NaCl layers can completely (or partially) desorb at much lower temperatures prior to becoming fully encapsulated by the GaAs. The direct growth of a monocrystalline GaAs layer directly on a NaCl thin film was only demonstrated recently by our group by using very short alternating pulses of Ga and As to promote adatom mobility, and better surface coverage with thinner layers.[25] The work discussed here will present the intricacies with the deposition of NaCl, subsequent GaAs deposition directly on rapidly desorbing NaCl thin films at elevated temperatures up to 500°C, and the convoluted interplay with exposure of the NaCl layer to a reflection high energy electron diffraction (RHEED) electron beam which is the foundation that led to the successful demonstration of single-crystal GaAs on NaCl.

## 2. RESULTS AND DISCUSSION

An array of intertwining results will be presented in the following section. First, the temperature dependence of the deposition of NaCl thin films on GaAs substrates will be presented in Section 2.1. The NaCl layer is crystalline and oriented parallel to the substrate with compositionally abrupt interfaces. With thin films of NaCl being reproducible, Section 2.2 pertains to work done on the growth of GaAs on these thin NaCl layers with a large focus on the substrate temperature. Initializing the GaAs growth at low temperature (<300°C) yields poor crystallinity of GaAs on NaCl thin films. However, NaCl begins to decompose rapidly above 300°C; withholding deposition of the GaAs overlayer until temperatures >500°C results in complete desorption of the NaCl layer, and homoepitaxial GaAs layers. The NaCl must be capped rapidly at lower temperatures to combat excessive desorption of the NaCl at elevated temperatures. However, GaAs growth on NaCl proceeds via the formation of discrete islands and the relatively thick layers that must be grown at low temperatures in order to coalesce largely spaced islands without desorption of the NaCl results in poor crystallinity.

Sections 2.3 and 2.4 will show how exposure of the NaCl surface to the RHEED beam can positively affect the nucleation of GaAs. We show that the presence of the RHEED beam during and even prior to GaAs deposition influences the morphology of the overlayer. Large changes were observed between areas exposed to RHEED during the GaAs deposition; a scheme was developed wherein the NaCl surface was exposed to the RHEED beam under an As-flux at low temperature prior to the GaAs deposition. Under these conditions, an amorphous As layer would condense on the NaCl surface only where the RHEED beam was exposed, and the entire sample could be covered relatively uniformly. The amorphous layer desorbs from the surface upon heating to the temperature where GaAs is subsequently grown. The GaAs nucleates more rapidly and better protects the NaCl in these RHEED exposed areas. To best utilize this effect, the GaAs growth was eventually separated into a nucleation step where a thin (<100 nm) GaAs layer is deposited at low temperature, and then heated to 580°C for further GaAs growth at more typical conditions. This procedure resulted in highly textured GaAs on complete NaCl thin films after continued growth at 580°C and is the basis of additional work outside the scope of this paper which yielded monocrystalline GaAs.

### 2.1. Deposition of NaCl thin films on GaAs (001) substrates

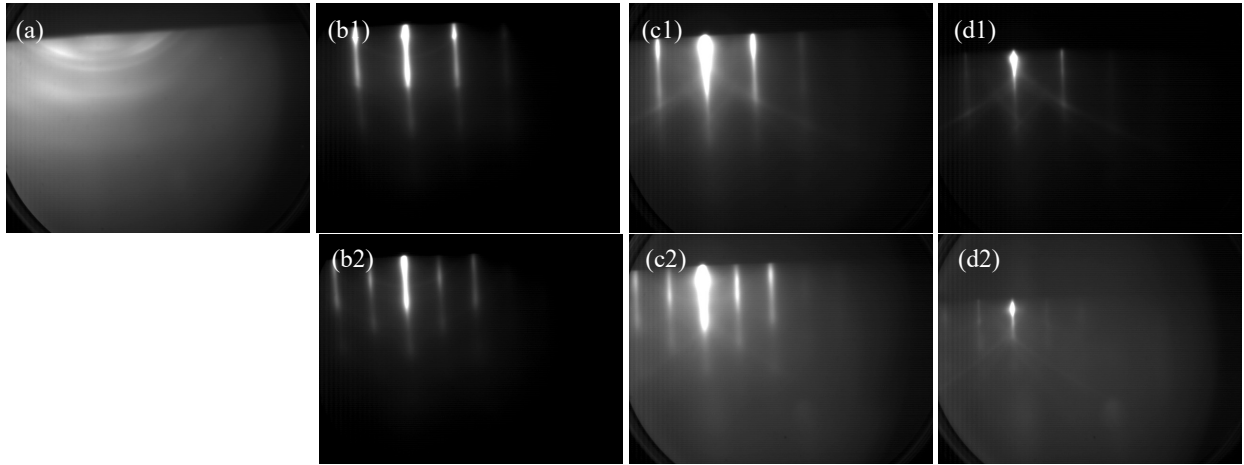

Figure 1: RHEED images of different samples with 10 minutes (roughly 30 nm) of NaCl deposition on (a) a GaAs substrate at 100°C with a small amount of excess As on the surface, and on clean GaAs surfaces at (b) 100°C, (c) 150°C, and (d) 175°C. Top row (1) taken along the [100]. Bottom row (2) taken along [110].

The first step is to understand the growth of crystalline NaCl on GaAs (001) substrates. First, a 300 nm GaAs buffer layer is grown at 580°C; after completion, RHEED reveals the typical As-terminated 2×4 surface reconstruction. The diffraction pattern converts to a symmetric c(4×4) as the sample is then cooled under an As-overpressure (~1×10$^{-6}$ Torr) until ~340°C. The sample is cooled to the desired temperature for the growth of NaCl ($T_{NaCl}$). A nominally 30-nm-thick layer of NaCl is deposited using a NaCl deposition time ($t_{NaCl}$) of 10 minutes and a growth rate of ~3 nm/min. If the As-source remains open when the substrate temperature is < 320°C, the RHEED pattern begins to go diffuse as a result of the condensation of small amounts of amorphous As. In this case, a ring-like pattern quickly appears upon opening of the NaCl shutter which does not revert back to a streaky single-crystalline pattern with further NaCl deposition (Figure 1(a1)). Blurry arcs superimposed on the rings signal some degree of fiber texturing,[26] which in the absence of a crystalline surface on which to nucleate, would suggest that NaCl has a preferred growth direction.

The remaining images of Figure 1 show RHEED patterns viewed along the [110] and [100] directions of 30 nm of NaCl deposited at various $T_{NaCl}$ starting on clean c(4×4) GaAs surfaces and a chamber pressure less than <7×10$^{-8}$ Torr. Deposition at 100°C (Figure 1(b)) is most closely lattice matched with GaAs and displays a streaky pattern with slight undulations indicating small island growth.[27] Figure 1(c) shows patterns from deposition of NaCl at 150°C. The streaks become brighter and Kikuchi patterns become visible indicating that the surface becomes increasingly well-organized and smooth despite a slight increase in the lattice mismatch (which increases from ~0% at 100°C to 2.9% at 580°C) due to the difference in thermal expansion between the materials. Figure 1(d) shows that deposition of NaCl at 175°C results in a dimmer pattern overall which could be for two reasons which will be discussed more in sections 2.2.2 and 2.3. First, the exposure of the NaCl to the RHEED beam begins to have substantial effects on the salt layer at higher temperatures. Second, as the substrate temperature is increased, the NaCl layer begins to decompose. Thus, the nucleation of NaCl at temperatures >175°C was not studied. When deposition occurs on clean GaAs surfaces (Figures 1(b-d)), the ratio of the spacing between streaks along the [110] and [100] directions is proportional to $\sqrt{2}/2$. Additionally, the [110] and [100] patterns of the NaCl are parallel to the [110] and [100] directions of the GaAs, indicating that the NaCl has a cubic symmetric crystalline surface.

Further insight into the crystallinity, structure, and diffusion of the NaCl layers was desired. Thus, 90 nm NaCl (deposited at 150°C) was capped with GaAs at 350-580°C (details on the GaAs deposition procedures will be discussed in Sections 2.2 and 2.3). Figure 2(a) shows a scanning transmission electron microscopy (STEM) image with fast Fourier transforms (FFTs) of both the NaCl and GaAs substrate. The FFTs reveal that the NaCl layer is oriented the same as the substrate and has nearly identical lattice constants. Organized atomic planes of NaCl are observed, but they are not perfectly straight. Contrast variations in the NaCl layer are artifacts resulting from damage induced by the electron beam during acquisition (further discussed in Section 2.3 and the Supplementary 1), which could be a reason for the tilting of some of the atomic planes. There is also a dark ~5 nm layer at the interface between the GaAs substrate and NaCl film. This layer appears amorphous, but the rings in the FFT of the GaAs area suggest that it might be nanocrystalline. It is possible that this is due to beam damage, but a separate investigation of the NaCl deposition on different GaAs surface reconstructions revealed some degree of textured polycrystalline growth during the early moments of deposition.[25] However, an organized (001) surface is achieved by the end of the NaCl deposition, as shown previously in Figure (1).

Figures 2(b,c) shows a lower magnification STEM image and the corresponding Energy dispersive X-ray spectroscopy (EDX) line profile. The image reveals the presence of a similar upper GaAs/NaCl interface but the EDX shows that the interface with the substrate is compositionally sharper than this upper interface. There is no Ga or As observed in the bulk of the NaCl layer. Additionally, the Cl is also well contained in the NaCl and no appreciable outward

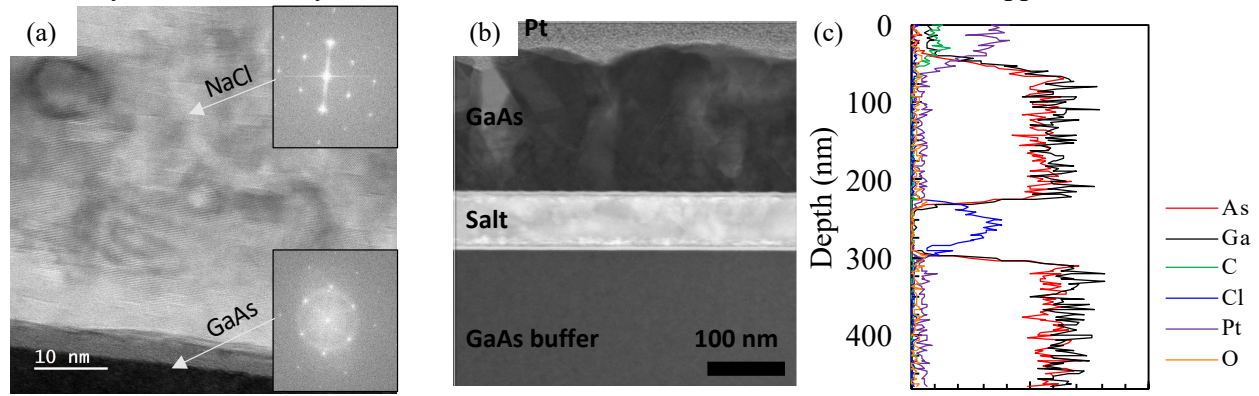

*Figure 2: (a) STEM image of the NaCl layer with insets showing FFTs of the (top) NaCl layer and (bottom) GaAs interfacial area (b) STEM image of a NaCl layer capped with GaAs, and (c) EDX maps of the NaCl layer contained between GaAs layers, and (d) line scans for each element.*

diffusion is seen. The signal for Na is not plotted here because the main K-peak overlaps the dominant Ga L-peak; the Ga signal is taken using the K-family peaks to avoid contamination from any Na signal. Thus, the Cl data combined with the TEM is used to verify the presence of a NaCl layer, and is only present within the bright layer. A slight increase in oxygen levels at both interfaces is observed, which is likely due to the growth pauses between the buffer layer and the NaCl, and the NaCl and the GaAs cap. This could also be related to the interfacial layers at both interfaces observed in the STEM image. The only presence of C is in the protective layer deposited during sample preparation, as also marked by the Pt signal.

## 2.2 Deposition of GaAs on NaCl thin films
### 2.2.1 Temperature dependence of initial GaAs nucleation

The subsequent deposition of GaAs layers on NaCl thin films was carried out in the same molecular beam epitaxy (MBE) chamber without any vacuum break. The first investigation involves varying the initial GaAs nucleation temperature ($T_{GaAs1}$) for GaAs deposition on a NaCl layer. A schematic of this growth process is given in the Supplementary 2(a). After growth of the high-temperature buffer layer, nominally 90 nm of NaCl is deposited at 110°C. The NaCl shutter is closed, and the temperature is increased at a rate of 50°C/min to the nucleation temperature under test. GaAs deposition begins once $T_{GaAs1}$ is reached at a rate of ~33 nm/min (calibrated using RHEED oscillations during homoepitaxial GaAs growth) by simultaneously opening both the Ga and As sources while the substrate temperature is continuously increased to 580°C. The resulting thicknesses of each sample are slightly different because the total GaAs deposition time ($t_{GaAs1}$) depends on the nucleation temperature and the ramp rate (samples with lower $T_{GaAs1}$ are thicker). Figure 3 shows RHEED patterns at the onset and at the end of GaAs deposition with corresponding cross section SEM images from a series of growths with various $T_{GaAs1}$. The additional spots in the RHEED patterns located to the left of the primary and first order spots here are artifacts of the incident RHEED beam, and not indicative of any surface reconstructions or twin grain structure.

When GaAs deposition begins at 100°C (Figure 3(a)) the RHEED pattern goes diffuse very quickly, signifying a lack of crystalline order. Due to the increased surface roughness and spontaneous delamination of the film from the substrate, the pattern gets darker upon continued growth and heating, as even diffuse reflections are blocked. In contrast to the other samples in this figure, the sample in Figure 3(a) has an additional ~90 nm of GaAs at $T_{GaAs1}$ (110°C) prior to heating to 580°C. Without this longer initial low-temperature deposition the film completely delaminated from the substrate during the growth. Cross sectional SEM reveals an extremely porous interface with a coalesced top layer. *Ex-situ* TEM measurements (not shown) reveal the smaller particles between the substrate and film to be crystalline GaAs, with the overlayer being a dense polycrystalline film with grains on the order of 100 nm. Additionally, if the temperature is not increased from 110°C, the film remains smooth, but fully amorphous and As-rich (Supplementary 3(a)). Additional samples (not shown) showed that changing the thickness of the initial 110°C deposition results in proportional changes in the thickness of the porous section. However, the coalesced top region remains similar in thickness, and without any observed improvements in the crystallinity. Thus, it can be assumed that the porous structure is a result of the low-temperature, As-rich, amorphous deposition and the coalesced layer is due to the growth at elevated temperatures.

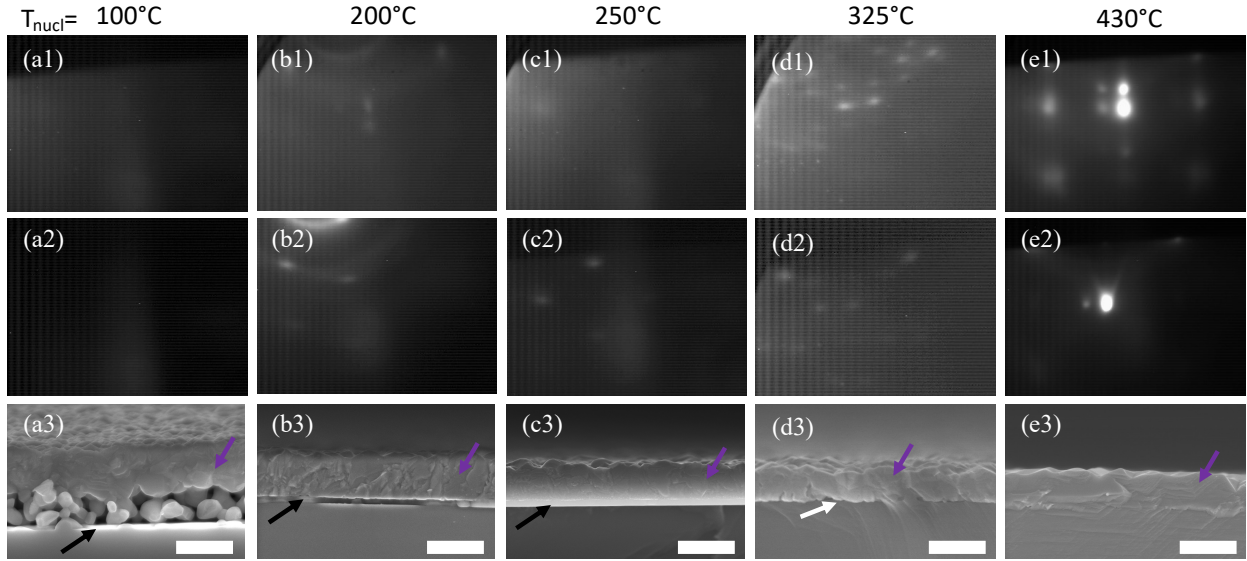

*Figure 3: RHEED images taken with a <100> beam direction during (1) the initial and (2) the final moments of GaAs deposition and (3) cross-sectional SEM of the finished sample (scale bar is 600 nm) after 30 minutes of salt deposition capped with GaAs initially at a temperature of (a) 100°C, (b) 200°C, (c) 250°C, (d) 325°C, and (e) 430°C and ramped to 580°C. Purple, black, and white arrows mark the GaAs overlayer, an NaCl layer, and any voids between the substrate and overlayer, respectively.*

The formation of a porous interface can be avoided by increasing the $T_{GaAs1}$ to 200-250°C (Figures 3(b,c)). In both cases the RHEED develops a spotty ring pattern upon initial GaAs deposition, which persists throughout growth indicating a textured polycrystalline film. The presence and persistence of the polycrystalline rings decreases as $T_{GaAs1}$ is increased. SEM shows the presence of a smooth ~70-nm-thick NaCl layer maintained beneath a fully dense ~500 nm GaAs overlayer. *Ex-situ* TEM and EBSD measurements reveal that these films are indeed polycrystalline, in agreement with the RHEED observations.

If the NaCl film is heated to temperatures ≥325°C prior to initiating the GaAs deposition (Figure 3(d-e)), the RHEED immediately begins to turn spotty, indicating Volmer-Weber style growth. Little change is observed throughout the deposition at 325°C. However, using a $T_{GaAs1}$=430°C, the pattern at the end of the GaAs deposition shows spots with chevrons indicating shallow surface facets. In either case with $T_{GaAs1}$ ≥325°C, SEM measurements no longer show the presence of a NaCl layer. Instead, large gaps between the substrate and a rough GaAs surface layer are observed. As the $T_{GaAs1}$ is increased, the large gaps become smaller pores, and eventually disappear altogether. The NaCl film begins to rapidly decompose as the sample is heated above 300°C; as a point of reference the NaCl effusion cell temperature is operating ~480°C.

The observation of the initial spotty RHEED in each scenario indicates that initial GaAs on NaCl growth proceeds via three-dimensional island growth. This results in incomplete coverage of the NaCl until all islands coalesce. This enables the NaCl to continuously desorb at the higher temperatures even during GaAs deposition. This happens especially quickly at higher temperatures where both the nucleation of islands is slower and the NaCl desorbs more rapidly. For example, at 300°C some GaAs is seeded on the NaCl before desorbing, but not quickly enough to form a cohesive film before the NaCl is completely desorbed, leaving behind large voids. By delaying GaAs deposition until 430°C, most of the NaCl has already desorbed from the surface. It is likely that by this point the NaCl layer is either very thin, or has completely disappeared, and many of the initial GaAs atoms are impinging directly on a GaAs surface, resulting in homoepitaxy. Any remaining NaCl escapes through pinholes or gaps between GaAs islands and the result is small pores in a mostly homoepitaxial structure. XRD and TEM measurements (not shown) confirm the RHEED measurements that the GaAs overlayers of any previously discussed case, with a persistent salt layer or large voids ($T_{GaAs1}$ ≤325°C), are polycrystalline. Additionally, using a $T_{GaAs1}$=580°C, the NaCl is completely desorbed prior to GaAs nucleation. In this case the RHEED remains very streaky, regains the typical reconstructions observed with MBE growth, and cross-sectional SEM shows no evidence that a NaCl layer was ever deposited. Intriguingly, this suggests that any NaCl can be thermally cleaned, and regrowth on substrates which once had salt deposited on them is possible.

### 2.2.2 Persistent NaCl layers at elevated temperatures

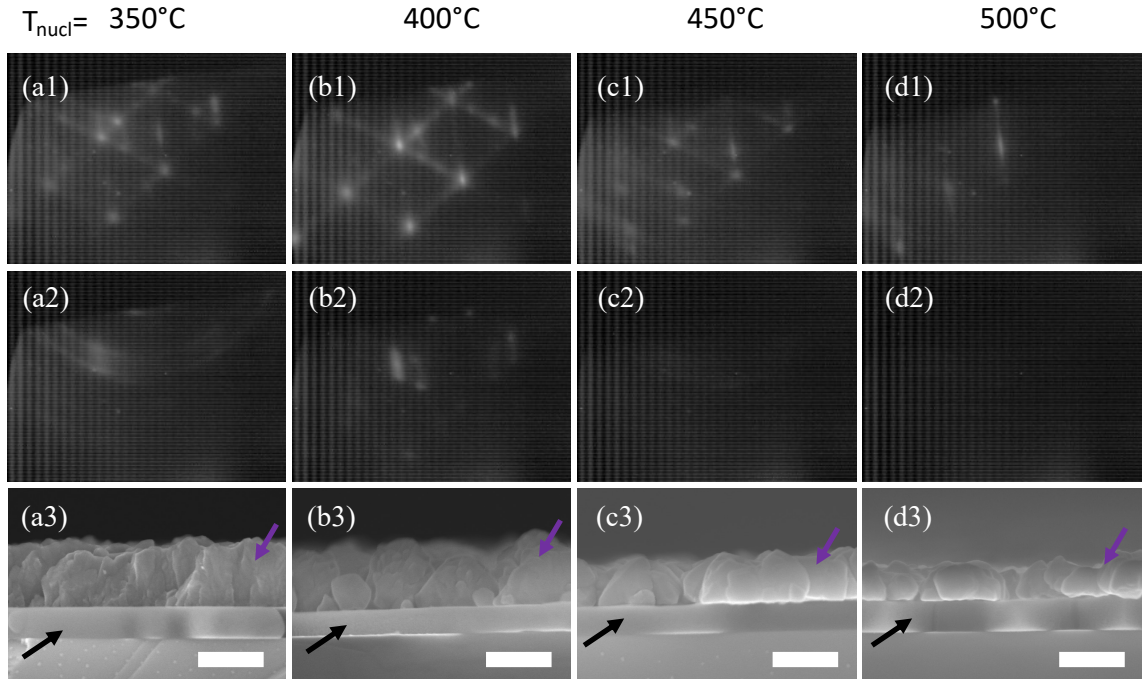

*Figure 4: RHEED images of (1) initial and (2) final GaAs deposition on thick NaCl and (3) the corresponding SEM images of samples with NaCl deposition beginning at 110°C and continuously until initializing the GaAs cap at (a) 350°C, (b) 400°C, c) 450°C, d) 500°C. Purple and black arrows mark the GaAs overlayer and NaCl layer, respectively. (scale bars are 300 nm)*

Investigation of GaAs deposited on NaCl at temperatures >300°C requires accounting for NaCl desorption upon heating. A new growth scheme was tested to combat this extra desorption where the temperature is now also held steady for the entirety of GaAs deposition ($t_{GaAs1}$=9 min) instead of being continuously ramped. An initial NaCl layer is deposited at 110°C ($t_{NaCl}$ up to 180 min) and is continuously deposited while increasing the temperature to $T_{GaAs1}$ at a rate of 20°C/min (outlined in the supplementary 2(b)). The time it takes to ramp to this temperature ($t_{ramp1}$) changes with the $T_{GaAs1}$ chosen. The highest temperature case ($T_{GaAs1}$=500°C) has a $t_{ramp1}$ ~19.5 minutes, for example. Additionally, the growth rate was increased up to ~17 nm/min. Thus, neglecting any desorption and accounting for the NaCl deposited during both $t_{ramp1}$ and $t_{NaCl}$ (19.5 and 180 min), the thickest NaCl layers deposited were ~3.3 μm. RHEED images of the NaCl surface during deposition and heating to the growth temperature are shown and discussed in supplementary 3.

RHEED and cross-section SEM patterns are used to analyze the growth of GaAs on the thick continuously-deposited NaCl films. Figures 4(a1-d1) show the RHEED patterns during the initial moments of GaAs deposition at various $T_{GaAs1}$. Samples with $T_{GaAs1} \leq 450°C$ show complex RHEED patterns with symmetric shadow spots and chevrons within the first ~10 s of deposition. The extra set of spots symmetric about the primary reflections are expected to be due to twins, likely rotations about the {111} which will be discussed more in Section 2.3.2. The shallow angle chevrons as viewed along this <110> direction correspond to {111} faceting of the GaAs. However, at $T_{GaAs1} = 500°C$ the RHEED during initial growth remained streaky and it took nearly 60 s to display a RHEED pattern similar to what was observed at the lower temperatures (Figure 4(d1)). In this case, it is possible that the initial GaAs islands are more epitaxially oriented and have a lower degree of twin formation. However, it is more likely that the longer time before a similar pattern is observed is due to a slower nucleation and growth of GaAs at this elevated temperature.

The middle row of Figure 4 shows the RHEED patterns at the end of the GaAs deposition at the different temperatures. For $T_{GaAs1} = 350°C$ (Figure 4(a2)), the pattern at the end of the ~300 nm deposition is spotted and ring-like, indicating a textured polycrystalline film, in contrast to the original pattern. Figure 4(b2) also shows a spotty ring pattern for the growth at $T_{GaAs1} = 400°C$. The rings are comparatively less pronounced than the spots, but the pattern is dimmer overall. This indicates a higher degree of crystalline order compared to growth at lower temperature, but still a very rough film. It was impossible to discern a pattern at the end of the growth for depositions at $T_{GaAs1} \geq 450°C$. The pattern become steadily darker as the growth temperature is increased, likely due to the formation of increasingly discrete islands. Electrons at RHEED energies have a mean free path on the order of a few tens of nanometers. Thus, these glancing-angle electrons are completely blocked by the islands which are hundreds of nanometers in size, and the pattern on the phosphor screen becomes dark.

The bottom row of Figure 4 shows the corresponding SEM images of the previously discussed samples. All samples have the same $t_{GaAs1}$ and growth rate (33 nm/min) and thus should have the same target thickness of 300 nm. Figure 4(a3) shows that GaAs deposited at 350°C is dense and approximately equal to the target thickness. As $T_{GaAs1}$ increases, the morphology trends toward the formation of more discrete faceted islands, in agreement with the RHEED observations from the early portions of the growth. Additionally, as the $T_{GaAs1}$ goes higher than 450°C the islands become smaller and more discrete. This also corresponds with the darkening of the RHEED pattern at these temperatures, because an array of disconnected ~200-nm-tall islands is essentially a very rough film. Figure 4(d3) shows that with $T_{GaAs1}=500°C$, the islands are thinnest (~190 nm), only ~60% of the targeted thickness. The reason behind this reduction in thickness, or apparent growth rate, with increasing temperature is not fully understood. It is possible that the rapid desorption of the NaCl surface at these higher temperatures creates chemical complexes, such as $GaCl_3$, which are more volatile and halt further growth or that impinging Ga and As atoms have difficulty remaining on such an actively desorbing surface, because the temperature of the substrate temperature is now higher than the NaCl effusion cell.

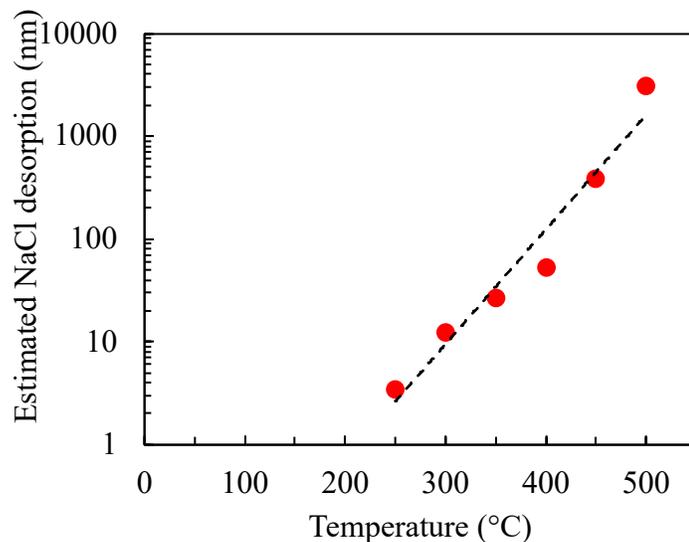

*Figure 5: Amount of salt desorption as a function of GaAs capping temperature*

Despite the NaCl thickness all appearing similar in these images, there are significant differences between the total amount of NaCl deposited. While the estimated and observed thickness of NaCl is similar when $T_{GaAs1} \leq 250°C$ (Supplementary 4), they begin to strongly diverge as the temperature increased. Figure 5 shows the estimated amount of total NaCl desorption using the difference between the expected thickness (from measured low temperature growth rates and total NaCl deposition time) less the thickness observed in the SEM images from Figure 4. As mentioned earlier, the thickest NaCl layer was deposited for $T_{GaAs1}=500°C$ (~3.3 µm), and the observed thickness is only ~160 nm. Thus, >3 µm of material had desorbed during the growth of this sample. The initial NaCl thickness required to maintain a persistent film until the end of growth increases exponentially as the growth temperature is increased. The actual desorption rate of the NaCl is temperature dependent, but lower and upper bounds on the desorption rate can be made by making two generous assumptions using the sample with $T_{GaAs1}=500°C$: the NaCl decomposition either occurs (1) over the entire 29 minutes at which the sample is >110°C or (2) exclusively during the 9 min growth at 500°C. These would result in an average desorption rate somewhere in the range of 112-362 nm/min. This desorption rate would be in the range of 3.4-11× the growth rate of GaAs used in this study. This presents an obvious challenge for achieving growth at typical GaAs deposition temperatures which are even higher.

## 2.3 The Effect of RHEED on GaAs/NaCl growth

### 2.3.1 Effect of RHEED exposure prior to and during GaAs deposition on NaCl

As demonstrated throughout the earlier sections, RHEED is a critical tool for *in-situ* observation, but the presence of the electron beam during growth actively affects the growth surface, which has been seen in other material systems.[28,29] In an attempt to elucidate the effects of the presence of the electron beam, a sample was grown where the beam was moved across the surface at different points during the growth, resulting in different levels of exposure and exposure starting at different times during the growth process. There is an obvious difference between regions with and without beam exposure even by eye. An image of the marks left by the 15 kV RHEED beam on this sample and discussion of the RHEED patterns from the growth are contained in Supplementary 5. The growth process used here was similar to that discussed in Section 2.2.2, i.e., 10 min of initial NaCl deposition at 110°C, then continuously deposited NaCl while heating to 300°C, at which point GaAs is deposited at a rate of 33 nm/min to reach a nominal thickness of 100 nm.

Figure 6 shows plan-view SEM images of seven distinct areas on the same sample with various degrees of RHEED exposure (RE) at different times throughout the growth at low and high magnification, top and bottom rows, respectively. Details, RHEED patterns, and a schematic of the RHEED exposure during the growth process are further discussed in Supplementary 5. The first column (Figures 6(a1, a2)) shows an area with no RHEED exposure (NRE) at any point throughout the growth. There are two different areas observed: a dark area and a light area. In this case the dark area is the NaCl and the light area is GaAs. This was inferred from the continuous degradation and movement of this surface under the presence of a tightly focused electron beam

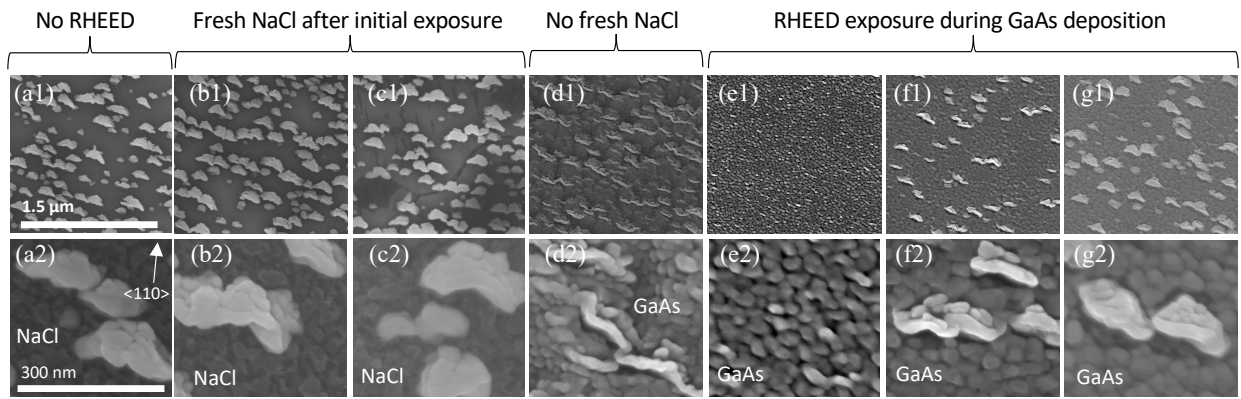

*Figure 6: Plan view SEM images of 100 nm of GaAs deposited on a NaCl film at 300°C at (1, top row) low and (2, bottom row) high magnification. (a) was not exposed to RHEED at any point during the growth. (b) was exposed to RHEED during NaCl deposition and covered with >20 nm of NaCl without RHEED exposure. (c) NaCl exposed to RHEED for 180s and covered with 4.5nm of NaCl without exposure. (d) NaCl exposed to RHEED for 90s immediately prior to GaAs deposition. RHEED exposed to the area during the (e) initial, (f) second, and (g) third minute of GaAs deposition*

in the SEM (details in Supplementary 1). The difference in smoothness of the dark surface can be seen between the low and high magnification images; it was impossible to acquire a high magnification image without some roughening of the NaCl layer. However, the light areas, do not degrade. The composition of these areas is not known, but they are highly ordered, with the long axis nearly perpendicular to the <110> direction (all images in Figure 6 are oriented the same).

The images from the next three columns show areas exposed to the RHEED beam prior to GaAs deposition. Figure 6(b) shows an area with RE during the initial low temperature NaCl deposition but eventually covered with ~20 nm of fresh NaCl at 300°C while the RHEED beam was moved to the next areas. This area looks very similar to the area with NRE; it seems like 20 nm of NaCl deposition negates any effect from previous RHEED exposure. The area in Figure 6(c) was exposed to the beam for 180 s, which was just enough to see the RHEED pattern go slightly spotty (Supplementary 5). Once the transition in the RHEED pattern was observed, the beam is moved to the next location, and 4.5 nm of NaCl was subsequently deposited on this location (in Figure 6(c)) prior to GaAs deposition. These images look mostly similar to the previous areas; there is a similar density of lighter islands contrasting against a dark NaCl background. However, the NaCl (dark) area in the low magnification image (Figure 6(c1) shows two slightly different contrast areas with cracks between. This is presumed to be a roughening of the NaCl layer due to the constant exposure of the RHEED beam at 300°C. Again, this is not seen in the area that was similarly exposed yet had ~20 nm of new NaCl deposition (Figure 6(b)). Thus, it appears that 5-20 nm of subsequent NaCl deposition can hide any effects of the RHEED on the original NaCl layer. It took 180 s of constant exposure (during continuous NaCl deposition) for a change to be seen in the RHEED at 300°C. To observe a similar change at lower temperatures takes significantly longer or is never seen. It is expected that because this significant difference is seen with 180 s of exposure, similar changes that may not be observable in SEM or RHEED are likely happening with shorter exposures and/or at lower temperatures.

The images in Figure 6(d) are different from all the others, supporting the hypothesis that there are effects with shorter exposures. This area was exposed to the RHEED beam for 90 s while continuously depositing NaCl (not long enough to see a change in the RHEED pattern) up until the final moment prior to GaAs deposition. Thus, this region has no fresh NaCl deposition, and is the reason for the additional 4.5 nm of NaCl in the area shown in Figure 6(c). Here, the morphology of both the light and dark regions are distinctly different. The lighter features are thinner than in the previous case, but still have similar directionality. The darker region shows some larger, scale-like undulations that could be related to a rougher underlying NaCl layer and no longer degrades under the presence of the SEM electron beam, similar to the next figures.

The location of the RHEED beam was moved to three different areas for the first, second, and third minute of GaAs deposition (Figures 6(e-g)). The dark background from each of these areas is similar and comparable to Figure 6(d) without the large-scale undulations. The rough island-like morphology is also observable at low and high magnifications, does not degrade under the presence of the RHEED beam at any high focus conditions, and looks similar to previous studies of stoichiometric GaAs.[30] Figure 6(e) shows the area with RE during the first minute; the presence of the light islands is completely suppressed. Figure 6(f) shows that exposure during the second minute results in light islands with a similar shape to that observed in the area with RE just prior to GaAs deposition (Figure 6(d)). However, the density of these islands is lower compared to areas with NRE. Figure 6(g) shows the area with RE during the final minute of GaAs deposition; the density of light islands is similar to regions that were never exposed at all or had fresh NaCl after exposure.

The composition and mechanism behind the formation of these light islands are still unknown and a subject of investigation. The RHEED observations (Supplementary 5) support 3-D textured polycrystalline growth in all regions from Figures 6(e-g). Based on the results presented, it seems that the presence of a RHEED beam, either immediately prior to or during GaAs deposition, facilitates the growth of the small dark islands in these regions (likely GaAs). The formation of the light islands seems to start immediately upon GaAs exposure, but begins to stagnate after the first minute. These could be similar to the discrete GaAs islands observed in Section 2.2. The presence of RHEED facilitates nucleation of dark GaAs islands, which suppresses the formation of the light islands. Electron bombardment of NaCl surfaces was seen to promote nucleation in metal films.[23] However, if RE is withheld until light islands have already formed (Figures 6(f,g)), the smaller darker GaAs islands grow around them, and suppress formation of further light islands. There is no observable density of the dark GaAs islands in areas with >4.5 nm of fresh NaCl. This suggests that <5 nm of fresh NaCl is enough to erase the effects of RHEED, and suppresses the nucleation of islands with this morphology at 300°C. The effects of the RHEED beam were not limited to the growth of binary material but were also seen in the deposition of Ge on NaCl films (Supplementary 6), where RE also enhanced the nucleation density of the Ge islands. The presence of an electron beam during material deposition results in a uniform surface morphology underneath the millimeter-wide beam. However, this would not be a practical way of achieving any large area uniformity because the nucleation of GaAs is highly sensitive to the time exposed to the RHEED beam, both total duration and the point during the deposition.

### 2.3.2 RHEED induced As adsorption at low temperature

Another growth scheme was developed utilizing RHEED exposure only prior to GaAs deposition to achieve uniform nucleation in areas larger than the millimeter-wide RHEED beam. A detailed description with RHEED images is outlined in Supplementary 7. Immediately after a NaCl layer is deposited at 150°C, an As-flux of 12.2 atoms/nm$^2$s (equivalent to the Ga flux used for the 33 nm/min growth rate) was supplied to the surface. The surface was then exposed to the RHEED beam, and the pattern slowly goes diffuse signifying the condensation of amorphous material. The preferential condensation of amorphous As onto a bare NaCl surface was only observed in the presence of the electron beam. Thus, the beam was manually stepped across the surface until a diffuse pattern was observed over most of the sample. The total time required for this As-soak ($t_{soak}$) was typically 3-5 minutes. The substrate was then heated at a rate of 20°C/min under constant As exposure. Around 320°C the RHEED begins to transition from diffuse to streaky as the As desorbs from the NaCl surface.

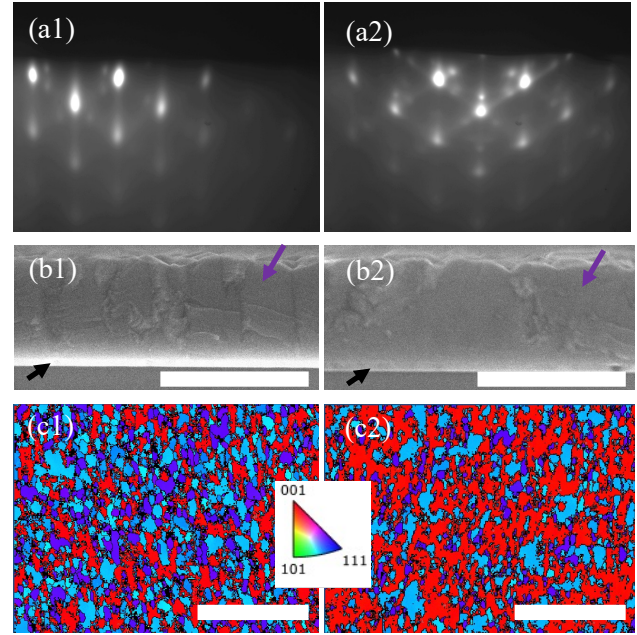

Figure 7: GaAs growth initialized at 350°C and ramped to 580°C (a) RHEED images, (b) SEM (scale bar 600 nm), and (c) EBSD maps of areas (scale bar 5 µm) (1) exposed to RHEED and (2) not exposed to RHEED until growth was completed. Purple and black arrows mark the GaAs overlayer and NaCl layer, respectively.

Figure 7 shows the RHEED patterns, SEM, and electron back-scatter diffraction (EBSD) maps of a sample utilizing this growth scheme. After ~30 nm of NaCl deposition, the RHEED was moved across the surface in the center portion of the sample under an As-flux prior to heating. Once heated to 350°C, GaAs growth was initialized and deposited while continuously heating to 580°C. The first column in Figure 7 shows the area with the RHEED assisted As-adsorption post-NaCl deposition. The second column shows the area with NRE. The RHEED patterns were taken at the end of growth and reveal slight chevrons in both cases. In the area with RE (Figure 7(a1)), the pattern is mostly spotty, with dimmer chevrons but slight streaking indicating a surface that isn't quite smooth and has a lower degree of any specific faceting. The dim secondary spots suggest some sort of twin-related secondary orientation. This contrasts with the pattern from the area with NRE (Figure 7(a2)) which was taken only at the end of growth to avoid any influence on the structure. Here, the chevrons are much brighter and shallow angle (similar to those observed in Figure 4), with brighter shadow spots, suggesting a higher degree of twin formation. The cross-section SEM images (Figures 7(b1,b2)) reveal similar thickness of both the GaAs and the underlying NaCl. Perhaps the roughness in the area with NRE could correlate with the spottier RHEED pattern as well.

However, plan view EBSD (Figures 7(c1,c2)) reveals significantly different crystalline textures between the two areas. There are a number of different orientations observed in the area with RE, where As condensed on the surface. A portion of the grains have the same {001} out of plane orientation as the substrate (red). The shades of purple and blue are results of different 90° azimuthal rotations of {223} grains and {122} grains, respectively. Quantitative pole figures (not shown) reveal that all {001} grains have the same azimuthal orientation as the substrate and that the prevalence of each 90° rotation of {223} is approximately equal. However, one of the rotations of the {122} grains is favored more than the other three. The EBSD data for the area with NRE shown in Figure 7(c2) reveals more area oriented commensurate to the substrate (red), and significantly fewer {223} grains (purple). A quantitative pole figure of this image shows that there is an extreme preference for the same single rotation of the {122} grain (blue) that was observed in the area with RE. Because the substrates used in this study have a maximum of 0.1° offcut, the distance between atomic steps would be at minimum 160 nm. Therefore, it is possible that there is some contribution from the step edges on the grains which are only hundreds of nanometers in size. The presence of these additional grains could also explain the shadow spots observed in the RHEED patterns at the end of growth. The root cause of the grain formation, as well as the single preferential orientation of these grains is the subject of an ongoing investigation.

From the data shown above, one could assume that the presence of RHEED was purely detrimental for the growth of near single crystal GaAs on NaCl. However, as temperature is increased the growth process with RHEED gets more complicated. Section 2.2 shows that the NaCl film is highly volatile and can completely desorb if not sufficiently capped before heating to elevated temperatures. A persistent NaCl layer must be maintained if one hopes to achieve liftoff of the overlayer. Section 2.3.1 shows that RHEED promotes the faster nucleation of GaAs, which would enable more rapid coalescence to protect the NaCl at higher temperatures.

### 2.3.3 Separate low-temperature nucleation and high-temperature growth

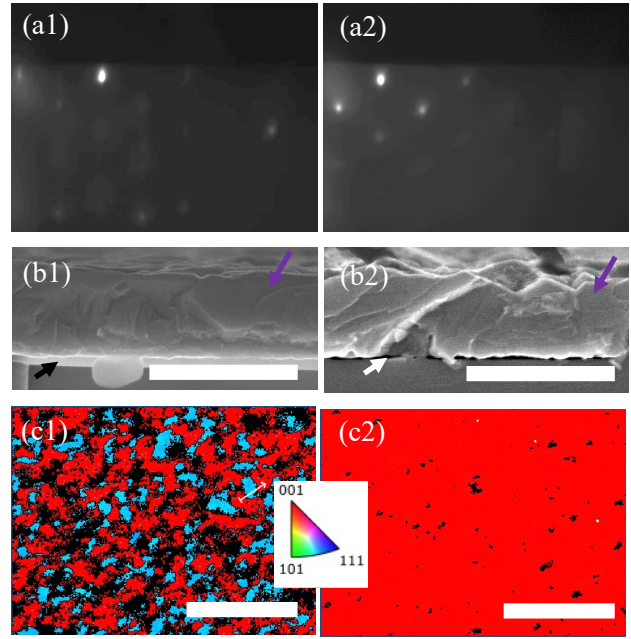

Figure 8: (a) RHEED images at the end of growth, (b) SEM images where purple, black, and white arrows mark the GaAs overlayer, NaCl layer, and voids between the substrate and overlayer, respectively (scale bars 600 nm), and (c) EBSD maps of areas (scale bars 5 µm) (1) exposed to RHEED and (2) not exposed to RHEED until growth was completed of a sample with a 3 minute GaAs nucleation at 400°C with additional 9 minutes of growth at 580°C.

The final growth scheme for this investigation was devised using everything shown until this point in an effort to achieve more crystalline GaAs overlayers (Supplementary 2(c)). After NaCl deposition at 150°C, the sample is exposed to an As-flux and the RHEED beam manually stepped across half of the sample surface. It was then heated to 400°C and ~100 nm of GaAs was deposited at a rate of 33 nm/min. However, now the growth is paused, and the sample is heated to 580°C. Once the sample reaches this temperature, 300 nm of GaAs is deposited.

The RHEED patterns from areas with RE and NRE are shown in Figures 8(a1,a2). The pattern from the area with RE is slightly streakier, while the pattern from the area with NRE shows faint steeper chevrons. This suggests that the area with NRE has a rougher faceted surface compared to the area with RE. Both areas are a bit dimmer but have substantially fewer additional spots than the samples in Section 2.3.2 that were grown continuously from the nucleation at lower temperatures. Figures 8(b1,b2) display SEM images which show a stark contrast between the two regions. A complete NaCl layer of the expected thickness is still present in the areas with RE. Conversely, in the area with NRE, there is no more NaCl and the overlayer has fused to the substrate. The area with NRE is unsurprisingly similar to samples grown at similar temperatures discussed in Section 2.2.1 (such as Figure 3 which showed fusion with the substrate when nucleating as cold as 325°C), while the area with RE is quite different. By utilizing the RHEED exposure at low temperature, GaAs was nucleated more quickly at 400°C, resulting in more rapid coverage of the NaCl, which was protected even during the high temperature deposition. Additionally, a relatively uniform area larger than just the width of the RHEED beam was able to be achieved because this RHEED sweep was done prior to any GaAs deposition (image of the sample in Supplementary 8(a)).

The EBSD maps in Figure 8(c1) show that there are only two predominant grain orientations in the RE area, those oriented commensurate to the substrate and a single orientation of the {122} (quantitative pole figure in Supplementary (8b). The {223} grains that were previously observed are now fully suppressed. The origin of this orientation must be a result of the lower GaAs nucleation temperature or the continuously depositing GaAs while heating. The EBSD from the area with NRE (Figure 8(c2)) shows monocrystalline GaAs. Because the material in this region does not show a complete NaCl layer, this could be a result from the bonding to the substrate, either homoepitaxy, recrystallization, or a combination of both. Without the presence of a NaCl layer it will not be possible to be simply removed from the substrate anyway. The increase in the amount of (100) overlayer material in the RE case is encouraging, because it indicates that formation of large area single crystalline material with persistent NaCl layers is achievable. Large area monocrystalline GaAs films on top of complete NaCl layers were realized[25] by carefully tuning additional parameters that were outside the scope of the work shown here, such as the GaAs growth rate, As/Ga ratio, and by utilizing techniques designed to promote adatom mobility.

### 2.3.4 Discussion of RHEED Effects

The RHEED effects of only three samples of GaAs on NaCl are shown in the previous section, but a large number of samples (>200) have been systematically grown and analyzed including some that deposit Ge directly on NaCl surfaces (Supplementary 6). They reveal that the effects of RHEED are quite complex, and neither exclusively beneficial nor detrimental. Some general observations and conclusions are outlined below:

*(1) The presence of RHEED roughens NaCl.* This effect is likely reduced with reducing the accelerating voltage, but then the pattern is too dim to be interpreted. This effect also seems to be more pronounced at higher temperatures i.e. at 15 kV it takes ~180s for the RHEED pattern to transition from streaky to spotty at 300°C, but at 150°C even after exposure for significantly longer times the RHEED is unchanged. Previous reports have suggested that the electron beam results in dissociation and desorption of the NaCl.[23,31] This could also be why imaging bare NaCl layers under highly focused electron beams in SEM and TEM results in the changes and destruction of the NaCl discussed in Supplementary 1. Unfortunately, most other *ex-situ* techniques to look at more subtle changes between areas with RE cannot be used on bare NaCl films because the NaCl degrades appreciably in the presence of water vapor, even in the amount present in the air.

*(2) RHEED and delamination.* The presence of the RHEED sometimes causes delamination. This is most frequently observed at the edges of the RHEED exposed areas. For samples with deposition temperatures >350°C, the area with RE shows a complete NaCl layer beneath a GaAs overlayer, while the area with NRE shows frequent fusion to the substrate. Delamination is somewhat obviously not observed in areas that have frequent fusion to the substrate, but it is also never observed in the regions that have a complete NaCl layer. However, at the transition between these two regions, the film is occasionally observed to be pulling away from the substrate; there is no NaCl and the fusion to the substrate is less frequent. For samples which use the As-sweep technique, this is thought to occur because this transition region only somewhat protects the NaCl due to less enhanced GaAs nucleation compared to areas with longer, more deliberate RE. Thus, this area has larger islands, which the NaCl sublimates out from underneath during prolonged high temperature steps, resulting in larger voids. This is especially true for films with thin nucleation layers heated to high temperatures for thicker (>300 nm) high temperature grown layers. This effect is also observed in the case with prolonged RHEED exposure. But in this case it could be the result of the RHEED roughening the film, discussed in the previous point, to the extent that holes are formed completely though the NaCl layer, which allows for less frequent fusion of the overlayer to the substrate and subsequently delaminates.

*(3) Excessively long RHEED exposure times do not enhance crystallinity of GaAs*. While some beneficial effects may be gleaned from selective exposure, prolonged exposure never seemed to give desirable results. This could be due to surface roughening or related to the higher degree of polycrystallinity observed in areas with RE. Constant exposure during the GaAs deposition resulted in a more polycrystalline material and sometimes spontaneous delamination of the film during growth.

*(4) The presence of RHEED enhances As-adsorption*. As discussed in section 2.3.2, arsenic preferentially condenses where the RHEED beam is present. The reason is not fully understood, but it is possible this is from a slightly rougher surface facilitating nucleation points, or from a change in surface charge promoting adhesion of As adatoms. As the temperature is increased, this amorphous As layer has to desorb before the NaCl surface atoms can, so in some way it can protect the NaCl at elevated temperatures. However, they both begin to desorb at similar temperatures, so the impact is limited. Related to the previous points, if the bare NaCl is exposed to the RHEED beam for too long the overlayer tends to delaminate from the substrate.

*(5) The presence of RHEED promotes nucleation of GaAs on NaCl*. This seems to be true not just during the actual nucleation step, but even exposure prior to opening the Ga shutter with the As adsorption. It is not known whether this is due to slight roughening of the NaCl or from the actual As-adsorption itself as it may not necessarily be fully desorbed by the time the Ga shutter is opened. However, this enhanced nucleation rate is one of the key benefits of RHEED for the formation of higher quality GaAs films on NaCl because swift formation of a complete GaAs layer is crucial to enable higher temperature depositions without sublimation of the NaCl layer.

*(6) RHEED affects the crystallinity of GaAs grown on NaCl using traditional co-deposition techniques.* In the cases discussed above it is possible that RHEED (or at least the As adsorption) is a cause of the highly textured GaAs overlayer because RE areas have more of the {223} grains than areas that are not exposed. For some samples such as the one where GaAs was deposited continuously from 250-580°C (Figure 3(c)), EBSD revealed that areas with RE showed grain sizes >5 μm, while the grains in areas with NRE were only a few hundred nanometers. However, there were no grains oriented commensurate to the underlying substrate or NaCl layer in either area. This could be related to previous studies on electron beam annealing of amorphous GaAs,[32,33] but these studies were done with much higher accelerating voltages.

## CONCLUSION

In conclusion, thin NaCl layers have been deposited on GaAs (001) substrates, as well as GaAs on top of the NaCl. The temperature at which the GaAs is nucleated has great effects on the crystallinity and overall morphology of the end film ranging from discrete islands, to extremely porous interfaces, to fully dense films with sharp interfaces. The presence of a RHEED beam, both prior to and during GaAs nucleation, can profoundly change the structure of the overlayer. This is in large part because of the enhanced nucleation of GaAs islands, enabling more rapid coalescence of a GaAs film to protect the volatile NaCl layer. Through combining careful RHEED exposure and separating low temperature nucleation and high temperature growth of the overlaying GaAs layer, highly textured GaAs films on persistent NaCl layers were achieved.

**METHODS**

An Epi930 MBE chamber is used to deposit NaCl layers on GaAs (001)±0.1° substrates. The GaAs substrate is heated to 620°C for 25 minutes under exposure to As before growth to ensure a clean and oxide free surface prior to deposition of a 300 nm GaAs buffer layer at 580°C. Ga is supplied from a standard effusion cell and As is provided by a valved cracker source. The substrate is then cooled under an As overpressure until ~340°C, after which the As is closed, and it is cooled to the temperature of NaCl deposition. NaCl is deposited via sublimation of 5N NaCl from a conventional effusion cell. Deposition temperatures investigated for the growth of a NaCl layer and the subsequent GaAs layer range from 100-350°C and 100-580°C, respectively. Temperature is measured via band-edge thermometry using a kSA BandiT system. Reflection high energy electron diffraction (RHEED) with an accelerating voltage of 15 kV is used to measure the surfaces during growth.

The growth morphology and epitaxial relationships of the different layers were investigated using various methods of electron microscopy. SEM was performed on a Hitachi S-4800 using accelerating voltages from 3-7 kV and a beam current of 5-7 nA. EBSD data was acquired with the sample tilted at 70° using an Oxford system with a Symmetry detector and CMOS sensor technology and an acquisition voltage of 20 kV. TEM imaging, electron diffraction patterns, and EDS were carried out with a JEOL 2100F/Oxford Instruments X-Max EDS at 200 kV. The GaAs substrate was tilted so that incident electrons are along <110>. Bright field TEM imaging was performed to show both overall layers and atomic structure of defects. Electron diffraction within an area of 100 nm in diameter was acquired to identify the local phases and crystalline orientation. To minimize damage from long electron beam dwell times during EDS map acquisitions, the average information across regions of interest was collected. EDS mapping was conducted with Aztec in ≤ 10 min. Subsequently the counts of the EDS map were summed along the horizontal line perpendicular to the growth direction to achieve line profiles of elemental distribution.

**ACKNOWLEDGEMENT**

This work was authored in part by the National Renewable Energy Laboratory, operated by Alliance for Sustainable Energy, LLC, for the U.S. Department of Energy (DOE) under Contract No. DE-AC36-08GO28308. Funding provided by Shell International Exploration and Production Inc., Houston, USA. The views expressed in the article do not necessarily represent the views of the DOE or the U.S. Government. The U.S. Government retains and the publisher, by accepting the article for publication, acknowledges that the U.S. Government retains a nonexclusive, paid-up, irrevocable, worldwide license to publish or reproduce the published form of this work, or allow others to do so, for U.S. Government purposes.

# REFERENCES

[1] K.A. Horowitz, T.W. Remo, B. Smith, and A.J. Ptak, Technical Report https://www.nrel.gov/docs/fy19osti/72103.pdf (2018).

[2] K.J. Schmieder, E.A. Armour, M.P. Lumb, M.K. Yakes, Z. Pulwin, J. Frantz, and R.J. Walters, 2017 IEEE 44th Photovoltaic Specialist Conference, PVSC 2017 **7**, 1 (2017).

[3] R. Lang, F. Habib, M. Dauelsberg, F. Dimroth, and D. Lackner, Journal of Crystal Growth **537**, 125601 (2020).

[4] K.L. Schulte, A. Braun, J. Simon, and A.J. Ptak, Applied Physics Letters **112**, 1 (2018).

[5] J. Simon, D. Young, and A. Ptak, 2014 IEEE 40th Photovoltaic Specialist Conference, PVSC 2014 538 (2014).

[6] W. Metaferia, K.L. Schulte, J. Simon, S. Johnston, and A.J. Ptak, Nature Communications **10**, 1 (2019).

[7] S.W. Bedell, D. Shahrjerdi, B. Hekmatshoar, K. Fogel, P.A. Lauro, J.A. Ott, N. Sosa, and D. Sadana, IEEE Journal of Photovoltaics **2**, 141 (2012).

[8] D. Shahrjerdi, S.W. Bedell, C. Ebert, C. Bayram, B. Hekmatshoar, K. Fogel, P. Lauro, M. Gaynes, T. Gokmen, J.A. Ott, and D.K. Sadana, Applied Physics Letters **100**, 12 (2012).

[9] C.A. Sweet, K.L. Schulte, J.D. Simon, M.A. Steiner, N. Jain, D.L. Young, A.J. Ptak, and C.E. Packard, Applied Physics Letters **108**, (2016).

[10] M. Konagai, M. Sugimoto, and K. Takahashi, Journal of Crystal Growth **45**, 277 (1978).

[11] E. Yablonovitch, T. Gmitter, J.P. Harbison, and R. Bhat, Applied Physics Letters **51**, 2222 (1987).

[12] G.J. Bauhuis, P. Mulder, E.J. Haverkamp, J.J. Schermer, E. Bongers, G. Oomen, W. Köstler, and G. Strobl, Progress in Photovoltaics: Research and Applications **18**, 155 (2010).

[13] A.N. Tiwari, A. Romeo, D. Baetzner, and H. Zogg, Progress in Photovoltaics: Research and Applications **9**, 211 (2001).

[14] D.K. Lee, S. Kim, S. Oh, J.Y. Choi, J.L. Lee, and H.K. Yu, Scientific Reports **7**, 1 (2017).

[15] S. Sharma, C.A. Favela, S. Sun, and V. Selvamanickam, Conference Record of the IEEE Photovoltaic Specialists Conference **2020-June**, 0744 (2020).

[16] Y. Nakamura, K. Saiki, and A. Koma, Journal of Vacuum Science & Technology A: Vacuum, Surfaces, and Films **10**, 321 (1992).

[17] A. Wangperawong, S.M. Herron, R.R. Runser, C. Hägglund, J.T. Tanskanen, H.B.R. Lee, B.M. Clemens, and S.F. Bent, Applied Physics Letters **103**, (2013).

[18] G. Shimaoka and S.C. Chang, Journal of Vacuum Science and Technology **9**, 235 (1972).

[19] SHIMAOKA G and CHANG SC, J Vac Sci Technol **8**, 243 (1971).

[20] R.F. Steinberg and D.M. Scruggs, Journal of Applied Physics **37**, 4586 (1966).

[21] T. Evans and A.J. Noreika, Philosophical Magazine **13**, 717 (1966).

[22] N.G. Dhere and N.R. Parikh, Annual Proceedings - Reliability Physics (Symposium) **1**, 149 (1980).

[23] F. Cosandey, Y. Komem, and C.L. Bauer, Thin Solid Films **59**, 165 (1979).

[24] A.J. Shuskus and M.E. Cowher, *Fabrication of Monocrystalline GaAs Solar Cells Utilizing NaCl Sacrificial Substrates* (1984).

[25] B.J. May, J.J. Kim, P. Walker, H.R. Moutinho, W.E. McMahon, A.J. Ptak, and D.L. Young, Journal of Crystal Growth (in-review).

[26] F. Tang, T. Parker, G.C. Wang, and T.M. Lu, Journal of Physics D: Applied Physics **40**, (2007).

[27] A. Ichimiya and P. Cohen, *Reflection High Energy Eelctron Diffraction* (Cambridge University Press, 2004).
[28] T.H. Myers, A.J. Ptak, B.L. VanMil, M. Moldovan, P.J. Treado, M.P. Nelson, J.M. Ribar, and C.T. Zugates, Journal of Vacuum Science & Technology B: Microelectronics and Nanometer Structures **18**, 2295 (2000).
[29] B.L. Vanmil, A.J. Ptak, N.C. Giles, T.H. Myers, P.J. Treado, M.P. Nelson, J.M. Ribar, and R.D. Smith, Journal of Electronic Materials **30**, 785 (2001).
[30] R.C. Myers, B.L. Sheu, A.W. Jackson, A.C. Gossard, P. Schiffer, N. Samarth, and D.D. Awschalom, Physical Review B - Condensed Matter and Materials Physics **74**, 1 (2006).
[31] A. Friedenberg and Y. Shapira, Surface Science **87**, 581 (1979).
[32] M.W. Bench, I.M. Robertson, and M.A. Kirk, MRS Proceedings **235**, 27 (1991).
[33] X. Yang, R. Wang, H. Yan, and Z. Zhang, Materials Science and Engineering B **49**, 5 (1997).

# Consideration of the Intricacies Inherent in Molecular Beam Epitaxy of the NaCl/GaAs System

Brelon J. May[1], Jae Jin Kim[2], Patrick Walker[1], William E. McMahon[1], Helio R. Moutinho[1], Aaron J. Ptak[1], David L. Young[1]

[1]National Renewable Energy Laboratory, Golden, USA
[2]Shell International Exploration and Production Inc., Houston, USA

**SUPPLMENTARY INFORMATION**

**S1. Discussion of the fragility of NaCl layers in air, in the presence of water, and under electron beams in the SEM**

**S2. Schematics of sample growths including shutter sequencing and temperatures**

**S3. Behavior of NaCl layers during high temperature ramp/anneal**

**S4. Exclusively low temperature growth of GaAs on NaCl (without anneal)**

**S5. Details on multi-RHEED exposure sample**

**S6. Deposition of Ge on NaCl thin films and RHEED effects**

**S7. Discussion of low temperature As-adsorption and RHEED images**

**S8. Quantitative pole figures**

## S1. Fragility of NaCl layers in air, in the presence of water, and under electron beams in the SEM

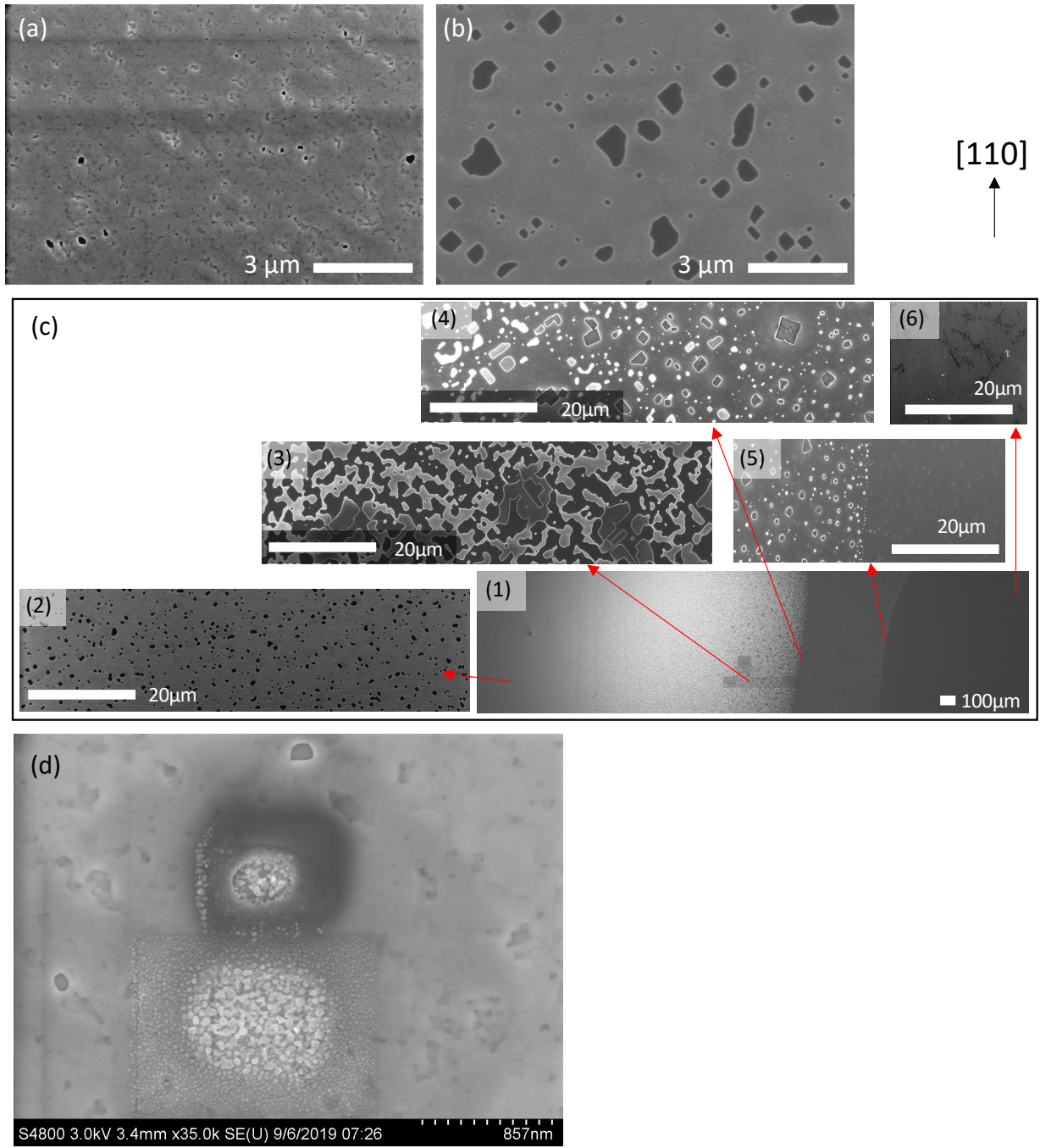

*Figure S1: Plan view SEM images of a bare NaCl thin film deposited on GaAs. (a) Image of the sample the day of growth and (b) 4 days after growth with most of the time in a nitrogen dry box. (c) Images of the sample the day of growth that was partially dipped in water for ~1s. (1) The low magnification image shows three distinct regions with the right side being the side dipped into water. Higher magnification images of the different regions (2-6) show lighter colored NaCl and the darker GaAs substrate. (d) Images of a NaCl surface with a 3kV beam at low magnification showing regions that were roughened form exposure at higher magnification*

The main manuscript discusses that the *in*-situ RHEED beam used for characterization has effects on the NaCl layer. However, bare NaCl layers are also extremely difficult to measure using *ex-situ* diagnostic tools. The {100} facets may be the low energy facet, and mostly stable in air, as one can see when using normal table salt. However, there are effects at small length scales that happen rapidly when removed from vacuum which has proved difficult for doing detailed analysis of bare NaCl layers. Figures S1 shows a variety of plan view SEM images of a 90 nm bare NaCl film deposited on a GaAs substrate. Figure S1(a) shows an image of the sample immediately after growth, and exposed to air while moving to the SEM. The surface is not as smooth as RHEED images would suggest. Some holes are also observed, it is not known whether these holes go all the way to the GaAs substrate, whether they form during growth or the walk over to the SEM. Figure S1(b) shows the same sample after storage in a nitrogen drybox for 4 days. The density and size of holes is now much larger. The sample is mounted with a cleaved {110} edge parallel to the sample holder, and these holes show very specific edges, which at ~45°, are likely the low energy {100} facets of NaCl. How these form in a dry environment is not a subject of this study, but rather here we focus on pointing out the difficulties in achieving consistent measurements between samples.

It is known that NaCl dissolves readily in water, but the extent to which thin films can be removed, and at what speed was tested here. The dissolution of NaCl was also investigated by partially submerging a piece of the bare NaCl (after storage in the drybox for 4 days). The sample was dipped in room temperature water by hand and immediately removed and blown dry with nitrogen before doing additional plan view SEM (Figure S1(c)). In the lowest magnification image (Figure S1(c1)) the right side of the image was dipped in water and the left was not; there are three clearly different contrast regions. Furthest from the water, in area (2), the NaCl surface still shows a large density of square holes resembling those seen in Figure S1(b). Moving further right, closer to the dipped area, the prevalence of the holes increases, and eventually the sample is only ~50% covered with salt (area (3)). Here, the NaCl still shows preferential directions with right angle faceting. Moving even closer to the edge (area (4)), one is left with discrete NaCl islands. These islands still have similar 90° edges and is likely right on the edge of where the water reached. Moving further right into the middle area, the islands no longer have such strong faceting or relationship to the substrate as seen in the left side of Area 5. This could be due to either the significantly higher vapor pressure of water at <1 mm from the surface in a dry environment (Denver, CO), instability during the water dip, or in the meniscus of the water surface, which could dissolve the NaCl, but then recrystallize as the highly NaCl-rich solution that was purely at the surface is dried. Beyond this line, nothing appearing like NaCl is observed suggesting that it is fully dissolved. There is some contrast/scratching on the GaAs surface but it is believed that a GaAs surface similar to the as-received case could be achieved by more carefully rinsing the surface with water, and subsequent thermal cleaning in vacuum and buffer layer growth. This all suggests that NaCl could provide a suitable release layer, but also study of bare layers in high humidity environments would be extremely difficult.

The final effect discussed in this section pertains to Figure S1(d) and is the fragility of the NaCl under exposure to electron beams. Here, another plan view SEM image is shown, taken with an accelerating voltage of 3 kV and beam current of 6.2 nA. Prior to acquisition of this low magnification image, the beam was focused in the two regions at a higher magnification (top) and medium magnification (bottom) which show significant effects from the electron beam. One can watch the surface move and evolve in real time while focused on these areas at higher magnification. Adjusting the focus and stigmation of the incoming beam i.e. oscillating the depth of focus, results in faster decay of the surface and deep holes can be made. It is not known whether this material is ablated, moved aside, or is different in crystallinity or composition than the original material. Additionally, as with the medium magnification area, using a long exposure time can result in degradation of the entire area. Thus, in order to acquire an image of a lesser damaged surface, one must move from the focused area to a new area and acquire the next image quickly. These effects are seen at accelerating voltages 3-5× lower than what is used in RHEED. It is possible that even though the electron beam in RHEED is at a glancing angle, rather than directly impinging on the NaCl surface, the degradation observed (such as observed in Figures S5(c,d)) could be because of a similar mechanism. The roughening of the surface observed via SEM could be comparable to what is happening when the surface is exposed to RHEED during the growth as well.

## S2. Schematics of sample growths including shutter sequencing and temperatures

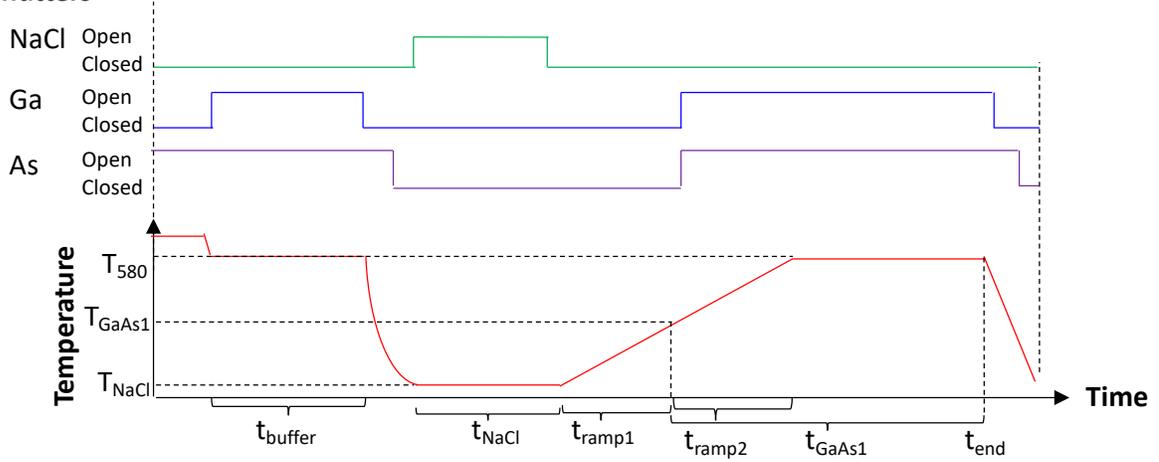

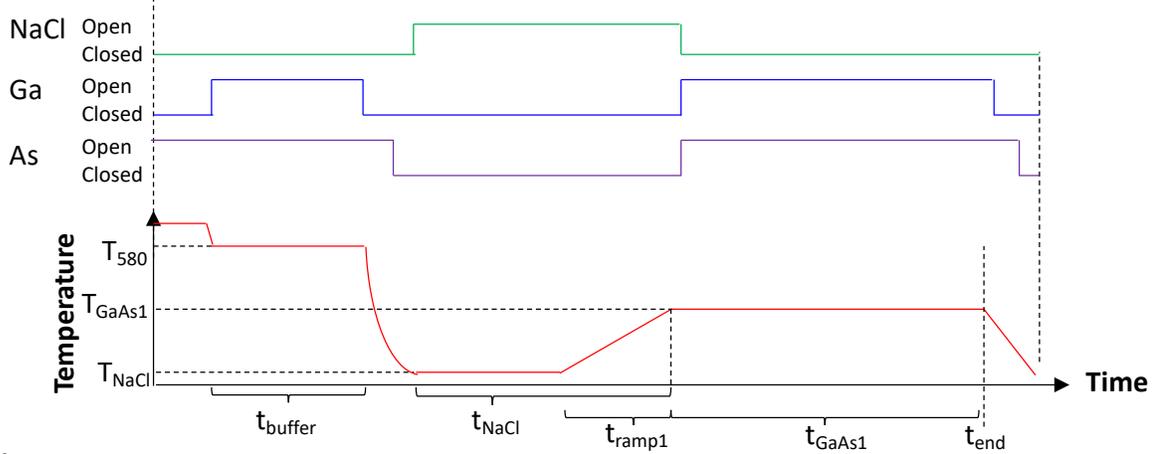

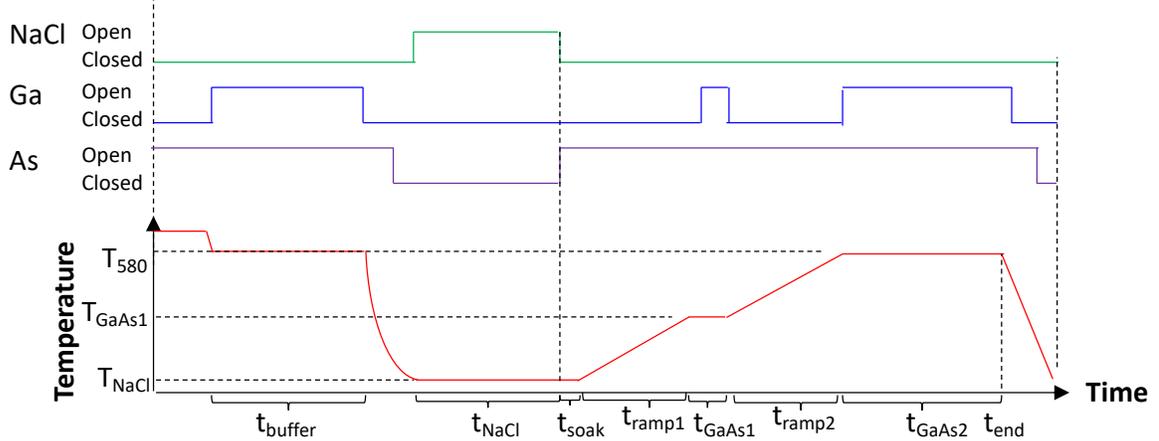

*Figure S2: Growth schematics showing shutter positions for NaCl (green), Ga (blue), and As (purple), with growth temperature (red) as a function of time for the growths discussed in (a) Figure 3, (b) Figure 4, and (c) Figure 8 in the main text.*

| Variable | Definition |
|---|---|
| $T_{580}$ | 580°C |
| $T_{NaCl}$ | Temperature of starting NaCl deposition |
| $T_{GaAs1}$ | Temperature at which the first GaAs deposition begins |
| $t_{buffer}$ | time of buffer layer GaAs deposition (9 minutes) |
| $t_{NaCl}$ | time of NaCl deposition |
| $t_{ramp1}$ | time taken to increase temperature from initial NaCl deposition temperature ($T_{NaCl}$) to GaAs nucleation temperature ($T_{GaAs1}$) |
| $t_{ramp2}$ | time taken to increase temperature from the GaAs nucleation temperature ($T_{GaAs1}$) to 580°C |
| $t_{GaAs1}$ | total time of first initial GaAs deposition |
| $t_{GaAs2}$ | total time of second GaAs deposition |
| $t_{soak}$ | time taken to sweep RHEED across NaCl surface under As-flux |
| $t_{end}$ | end of growth |

*Table S2: Definition of variables outlined in Figure S2*

Figure S2 shows schematics of the processes for the GaAs-capped growths discussed in the main manuscript. The shutter sequences for NaCl (green), Ga (blue), and As (purple) are shown in conjunction with the temperature of the substrate (red), as a function of time beginning from the oxide desorb step at 620°C, where the As shutter is already open. All samples go through this identical step, as well as the same time for a buffer layer ($t_{buffer}$) of 9 minutes to achieve a thickness of ~300 nm at 580°C ($T_{580}$), prior to cooling rapidly under an As flux until ~340°C. The samples are all then cooled until the temperature at which NaCl deposition begins ($T_{NaCl}$). The salt deposition time ($t_{NaCl}$) is varied between studies, and sometimes persists into the time taken to ramp temperature ($t_{ramp1}$) to the GaAs nucleation temperature ($T_{GaAs1}$). The substrate temperature ramp rate post-NaCl deposition is either 20 or 50°C/min depending on the sample set.

Figure S2(a) represents the growth scheme for Figure 3 in the main text. The NaCl is deposited only at low temperature. For these growths the ramp rate after NaCl deposition is 50°C/min. Thus, the time to ramp from $T_{NaCl}$ to $T_{GaAs1}$ ($t_{ramp1}$) and from $T_{GaAs1}$ to $T_{580}$ ($t_{ramp2}$) are both dependent on the $T_{GaAs1}$ chosen; $t_{ramp1}+t_{ramp2} \approx (580°C-T_{NaCl})/$(ramp rate). However, this is only approximately equal because it can take up to 5 minutes before the heater actually increases the substrate temperature. Additionally, the GaAs deposition time ($t_{GaAs1}$) is also dependent on $T_{GaAs1}$ as it includes $t_{ramp2}$ and the time of growth at 580°C. However, the time at 580°C is lengthened/shortened in an effort to keep the total thickness of the layers the same between sample comparisons, i.e. $t_{GaAs}$ is attempted to be kept the same for all samples. For these samples, once $T_{GaAs1}$ is reached, both the As and Ga shutters are opened simultaneously. The Ga shutter remains open for $t_{GaAs1}$, before closing and cooling the sample at the end of growth ($t_{end}$). The sample cools under an As-overpressure until ~340°C when the As shutter is also closed.

Figure S2(b) is a somewhat simpler scheme pertaining to Figure 4. The first part of the growth is the same as in Figure S2(a). However, in this case $t_{NaCl}$ also includes $t_{ramp1}$ as the NaCl shutter is kept open while ramping the substrate temperature, and a ramp rate of 20°C/min is used.

Thus, the thickness of the NaCl deposited is necessarily dependent on both $t_{NaCl}$ and the $T_{GaAs1}$ chosen. The temperature stops increasing once $T_{GaAs1}$ is reached. At this point the NaCl shutter is closed, and both Ga and As are immediately opened. The Ga shutter remains open for $t_{GaAs1}$. Once the growth is completed, the Ga shutter is closed. If $T_{GaAs1}>340°C$, the sample is cooled to ~340°C prior to closing the As shutter. If $T_{GaAs1}<340°C$, the Ga and As shutters are closed simultaneously.

Figure S2(c) pertains to Figure 8 in the main text and includes a two-step GaAs deposition and a time where arsenic is exposed to the NaCl surface. The time it takes to move the RHEED beam manually across the surface to effectively dim the diffraction pattern is $t_{soak}$. During this time, the As shutter is opened while the sample is still at $T_{NaCl}$. The As shutter remains open using an As flux with a As/Ga ratio = 1 (calibrated at 580°C) while heating the sample to $T_{GaAs1}$ at 20°C/min (which takes a predictable amount of time ($t_{ramp1}$). Once the $T_{GaAs1}$ is reached, the temperature stabilizes, and then the Ga shutter is opened for a short time ($t_{GaAs1}$) to form a thin layer at this lower temperature. The As shutter remains open as it is heated to $T_{580}$ using the same ramp rate (which takes $t_{ramp2}$) at which point the As flux is increased and the Ga shutter is opened for a secondary GaAs deposition time ($t_{GaAs2}$). Once complete, the same cooling procedure as outlined earlier is then applied.

This scheme can also be applied to Figure 7 in the main text (and Supplementary Figure S7). In this case there would be no growth pause, so $t_{GaAs1}=0$, and the Ga shutter would be open for the entirety of $t_{ramp2}$.

## S3. Behavior of NaCl layers during high temperature ramp/anneal

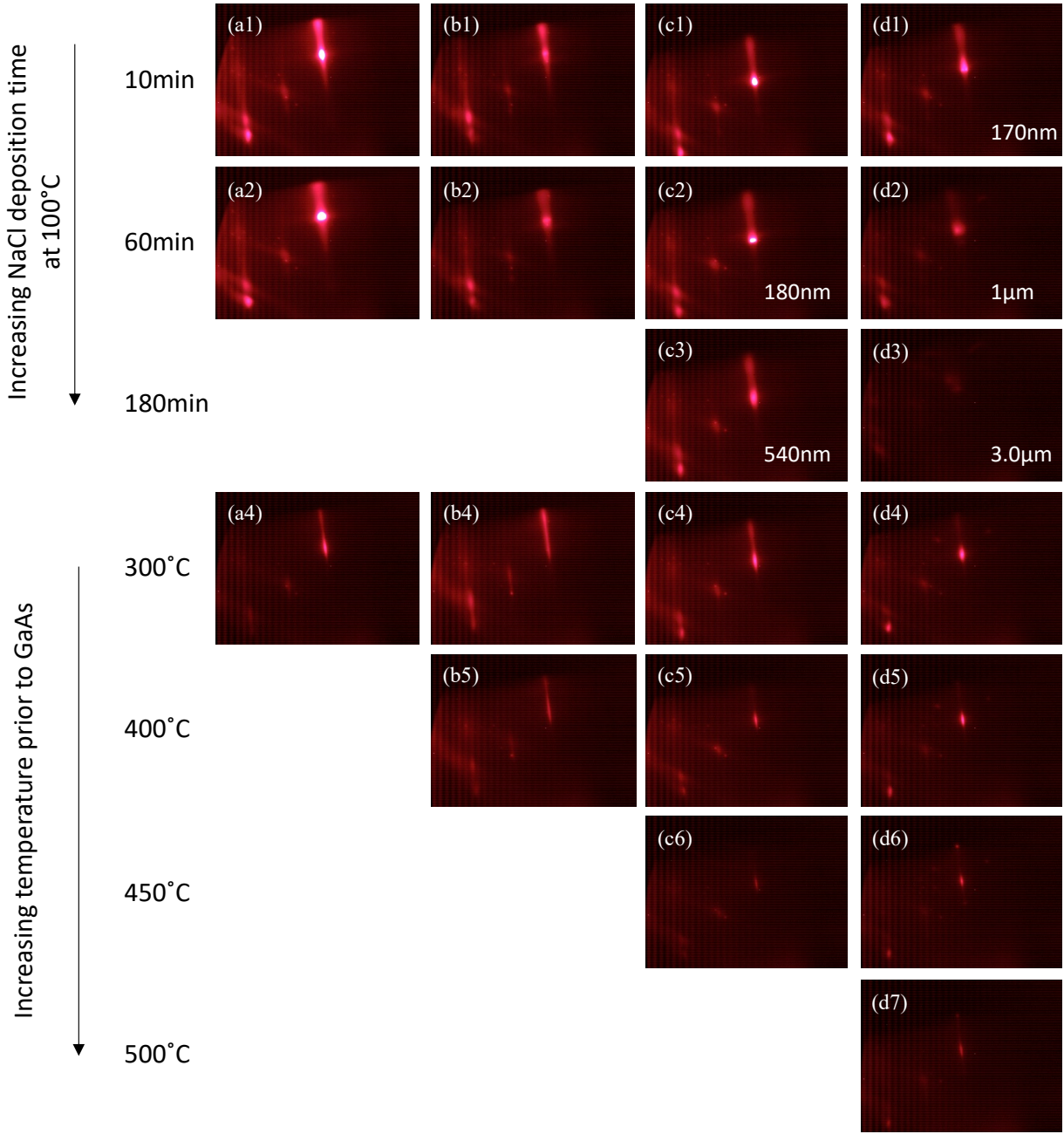

*Figure S3: Additional RHEED patterns along the <110> for samples discussed in Figure 4 in the main text consisting of NaCl deposited for various times at 100°C and capped with GaAs at various temperatures. (a) 72 min of salt capped at 350°C, (b) 74.5 min of salt capped at 400°C, (c) 197 min of salt capped at 450°C, (d) 199.5 min of salt capped at 500°C. RHEED images taken at (1) 10 min, (2) 60 min, and (3) 180 min into NaCl deposition at 110°C (where applicable) and of the NaCl surface with continuous NaCl deposition upon heating to (4) 300°C, (5) 400°C, (6) 450°C, and (7) 500°C.*

This section shows RHEED patterns taken prior to those shown in Figure 4 in the main text. As discussed in Supplementary section 2, for this growth scheme the thickness of the NaCl changes as it is based on $t_{ramp1}$. Additionally, thicker NaCl was required to maintain a layer at high temperatures. For growth temperatures ($T_{GaAs1}$) up to 450°C (Figures S4(a-c)), this was achieved by simply growing thicker NaCl layers using a growth rate of ~3 nm/min at 110°C (corresponding to a beam equivalent pressure (BEP) of ~7.3e-8 torr) and increasing the deposition time of NaCl ($t_{NaCl}$). To achieve a persistent NaCl layer with growth of GaAs at 500°C in a reasonable amount of time the growth rate of the NaCl was increased, the BEP of NaCl was increased to 4e-7 torr and NaCl was deposited for 180 min at 110°C (plus a $t_{ramp1}$ ~19.5 minutes for increasing the temperature from 110-500°C). Neglecting any desorption, this would be equivalent to growth of a layer that is ~3.3 µm thick.

Looking at the RHEED patterns for the lower growth rate samples, the patterns after 10 min (Figure S4(a1-c1) are the same as after 60 min of deposition (Figures S4(a2-c2)). For the case of higher NaCl growth rate (Figures S4(d1-d2)) the pattern dims slightly, although at this point there is already >1µm of material, and even at a lower growth rate, after 540 nm of material (Figure S4(c3)) the pattern started to dim. So it is possible that this dimming is only related to total thickness of the NaCl and not the higher growth rate.

For the lower growth rate samples, heating to 300°C and beyond shows a dimming of the RHEED pattern, which continues until $T_{GaAs1}$ is reached, and the GaAs capping begins. In the case of higher deposition rate required for $T_{GaAs1}$=500°C the RHEED pattern is slightly dimmer during the growth. However, the pattern begins to brighten as the temperature is increased to 300°C (Figure S4(d4)). This could possibly be attributed to a slight annealing of the layer at temperatures <300°C before the NaCl layer starts to desorb. But similar to the previous cases, the RHEED pattern begins to dim at >300°C, likely due to the increasing desorption and decomposition of the NaCl. In all cases there is at least a weak NaCl pattern still observable prior to beginning GaAs nucleation.

## S4. Exclusively low temperature growth of GaAs on NaCl

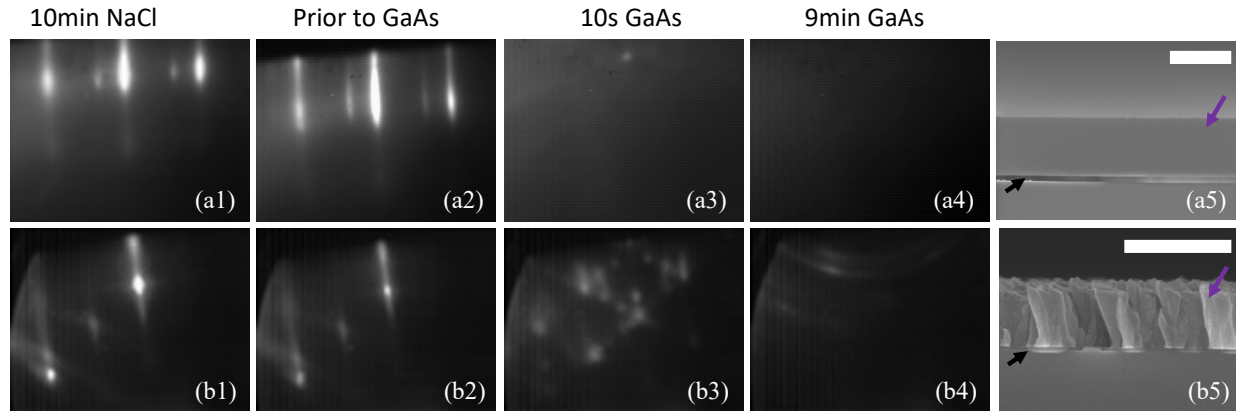

*Figure S4: (Top) RHEED images taken along a <100> direction (a1) after 10 min and (a2) 30 min of NaCl deposition at 110°C, and (a3) after 10 s and (a4) 9 min, of GaAs deposition at 110°C with (a5) a corresponding cross sectional SEM image of the growth.*
*(Bottom) RHEED images taken along a <110> direction (b1) after 10 min of NaCl deposition at 100°C and (b2) the NaCl surface once heated to 250°C just prior to GaAs deposition. Additional patterns (b3) after 10 s, and (b4) after 9 min of GaAs deposition with (b5) a corresponding cross sectional SEM image of the growth. Scale bars in both SEM images are 600 nm.*
*Purple and black arrows show the GaAs overlayer and NaCl layer, respectively.*

Some samples similar to those shown in Figures 3 and 4 in the main text were grown to look into the structure of purely low-temperature deposited GaAs on NaCl. These sample have no heating during the GaAs deposition, similar to Figure 4, but do not have continuous NaCl deposition. The RHEED again suggests that the NaCl layer is smooth, near single crystalline, and aligned with the substrate in both cases.

For the case of all 110°C deposition, the secondary spots in the RHEED patterns to the left of the peaks in the top row are not secondary phases or reconstructions and are due to the incoming beam having two primary spots two during this growth. The pattern from the NaCl surface quickly disappears upon opening of the Ga and As shutters and the pattern remains diffuse throughout the entirety of the growth. However, unlike the sample shown in Figure 3(a) in the main text, because this sample was never heated, a dense film with sharp interfaces is observed (Figure S3(a5)). TEM and EDS measurements (not shown) reveal that this deposited material is completely amorphous and As-rich.

The RHEED patterns in Figure S3(b) remain relatively unchanged when heating the NaCl to 250°C. However, without the continuous heating, the RHEED transitions from the streaky NaCl surface to showing many spots and chevrons, similar to patterns observed with higher temperature depositions in Figure 4 from the main text. First, the transition to a spotted RHEED indicates the formation of a three-dimensional surface. The chevrons passing through the first order spots viewed along [110] suggest GaAs islands take on a pyramidal shape with {111} facets. Extra spots symmetric about the primary and first order reflections also appear very early on in the GaAs deposition indicating the presence of twinning along the <111> directions. As the growth continues, this pattern fades and is replaced with rings with diffuse spots (Figure S3(b4)) similar to that observed during crystallographically textured nanowire formation.[26] SEM of this sample (Figure S3(b5)) shows complete coverage of a NaCl film with material consisting of densely packed but discrete columnar grains. This is in stark contrast to Figure 3(c) in the main text where growth is initialized at the same temperature, but continuously heated to 580°C.

## S5. Details on multi-RHEED exposure sample

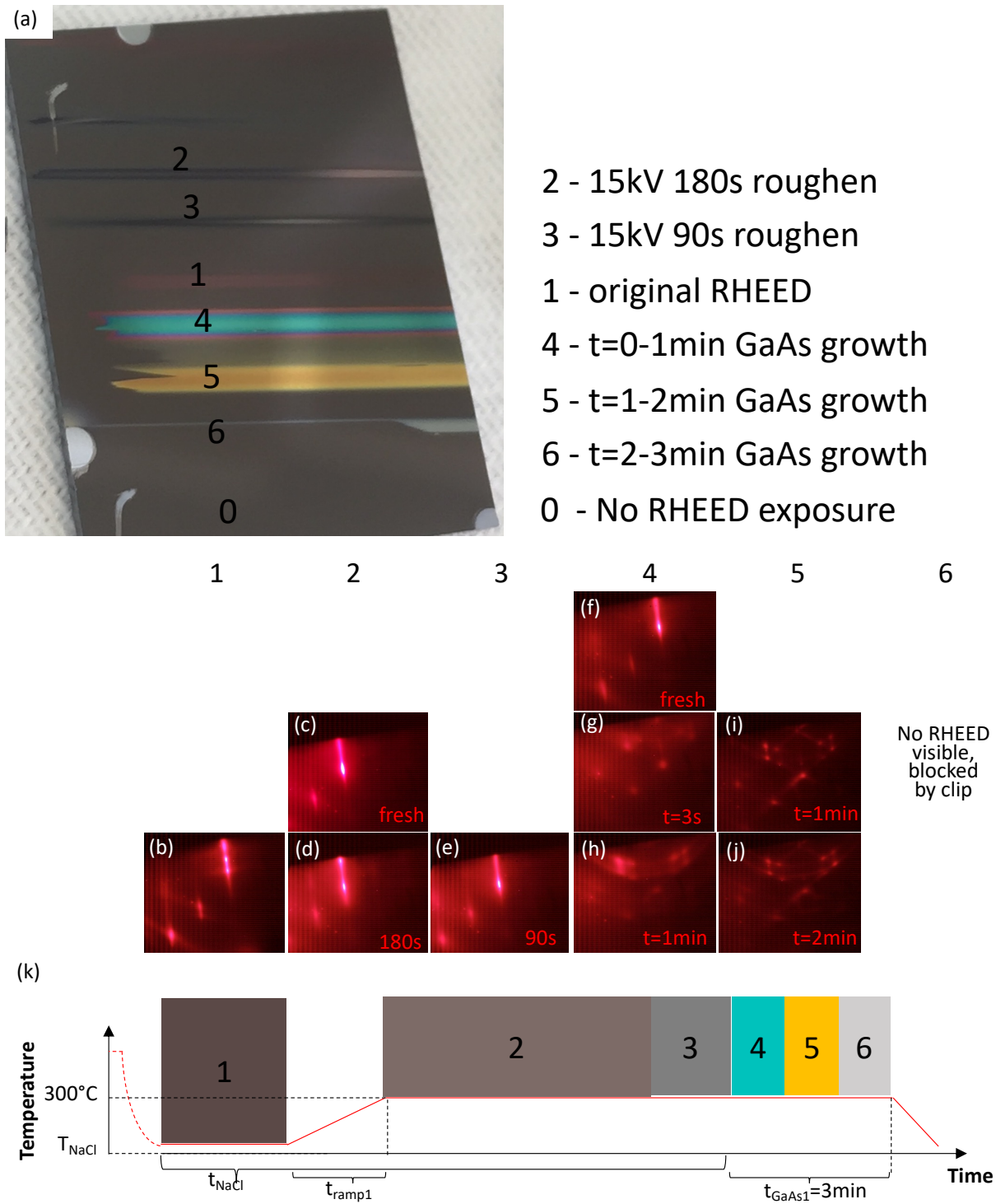

2 - 15kV 180s roughen
3 - 15kV 90s roughen
1 - original RHEED
4 - t=0-1min GaAs growth
5 - t=1-2min GaAs growth
6 - t=2-3min GaAs growth
0 - No RHEED exposure

*Figure S5: (a) Image of a sample after removal of the chamber showing distinct lines from exposure to the RHEED beam during different portions of the growth (labeled 0-6). (b-j) RHEED patterns from the different stages of growth in regions 0-6. (k) A representation of timing of the RHEED exposure for areas 1-6 from (a).*

Figure S5 contains details of the sample from Figure 6 in the main text, consisting of a 3 min GaAs deposition on NaCl at 300°C. A picture of the sample (Figure S5(a)) shows distinct color differences from the presence of the 15 kV RHEED beam at different portions of the growth. In the picture of the sample, areas exposed prior to GaAs deposition (lines 1 and areas above) appear relatively similar, while the three regions at different points during the GaAs growth (4, 5, and 6) have drastically different colors. These also presented morphological differences in plan-view SEM, which enabled acquisition of images with a high degree of spatial accuracy for Figure 6.

Area 0 has no RHEED pattern because it was never exposed, and Area 6 was blocked by the clip so no pattern could be obtained, so these will not be discussed. However, for Area 6 there was a region near the edge of the sample where the clip no longer shadowed the beam that was used for the plan-view SEM discussed in the main text. Figure S5(b) shows the pattern after 10 min of NaCl deposition at 110°C, it appears streaky as was observed in other growths. At this point the RHEED was moved to the edge of the sample (top of sample in Figure S5(a)) and turned off while the sample was heated (with continuous NaCl exposure). Figure S5(c) shows the brightened pattern of a fresh area of NaCl (during deposition) once the temperature reached 300°C. The beam is left on and unmoved while constantly monitoring for any change. Under continuous NaCl growth, it takes ~3 minutes for the RHEED pattern to begin to go spotty as shown in Figure S5(d) signifying that the surface was roughened. At this point the beam was moved to position 3 where it is left on for 90s. The pattern after holding the beam at Area 3 for 90s is Figure S5(e), and unsurprisingly looks like it would fall between Figure S5(c) and S5(d).

GaAs deposition began immediately once the beam was moved to the fresh NaCl surface in Area 4 (Figure S5(f)), which at first looks bright and streaky. The pattern after 3s of GaAs deposition (Figure S5(g)) shows the formation of new dim spots, with faint chevrons and shadow spots. After 1 minute (Figure S5(h)) the pattern has transitioned to spotty rings. This signifies that the growth starts out oriented but three dimensional, and after 1 minute of growth with continuous RHEED exposure ends up as a textured polycrystalline film. The beam is then moved to area 5 (Figure S5(i)) which does not show any ring-like characteristics. Rather, a spotty pattern similar to the onset of GaAs deposition with RHEED exposure (Figure S5(g)) is observed. However, the pattern is much less blurred, and the spots and chevrons are seen clearly. After a minute of continues deposition with exposure, rings start to appear again. This signifies that upon further deposition with RHEED exposure the growth starts to go more polycrystalline. Without the presence of the RHEED beam during nucleation, the clearer pattern observed suggests that the RHEED may be worsening the crystallinity of the nucleation layer. However, even without constant exposure to the RHEED beam, further deposition at 300°C would be polycrystalline, similar to Figure 4(a2) in the main text which was a longer growth performed at 350°C.

Figure S5(k) outlines the portion interest for the growth of this sample, neglecting oxide desorb and buffer layer growth steps, while highlighting the timing of the RHEED exposure in relation to the NaCl and GaAs deposition. The time axis is roughly to scale. The RHEED was not open for the entity of the time marked by region 1, and was turned on and off throughout this portion. However, for the remaining sections the RHEED was left on without any beam blanking. Note that NaCl is deposited for a considerable time (~270 s) at 300°C, $t_{NaCl}$ spans beyond the temperature increase, contrary to previous growths. This is the reason for further NaCl deposition on top of areas 1 and 2. After the 90 s roughening that occurs for area 3, the NaCl shutter is closed,

RHEED moved to area 4, and GaAs deposition initialized. The RHEED stays in each of the following 3 locations for 1 min before moving to the next without any pause in GaAs deposition.

## S6. Deposition of Ge on NaCl thin films and RHEED effects

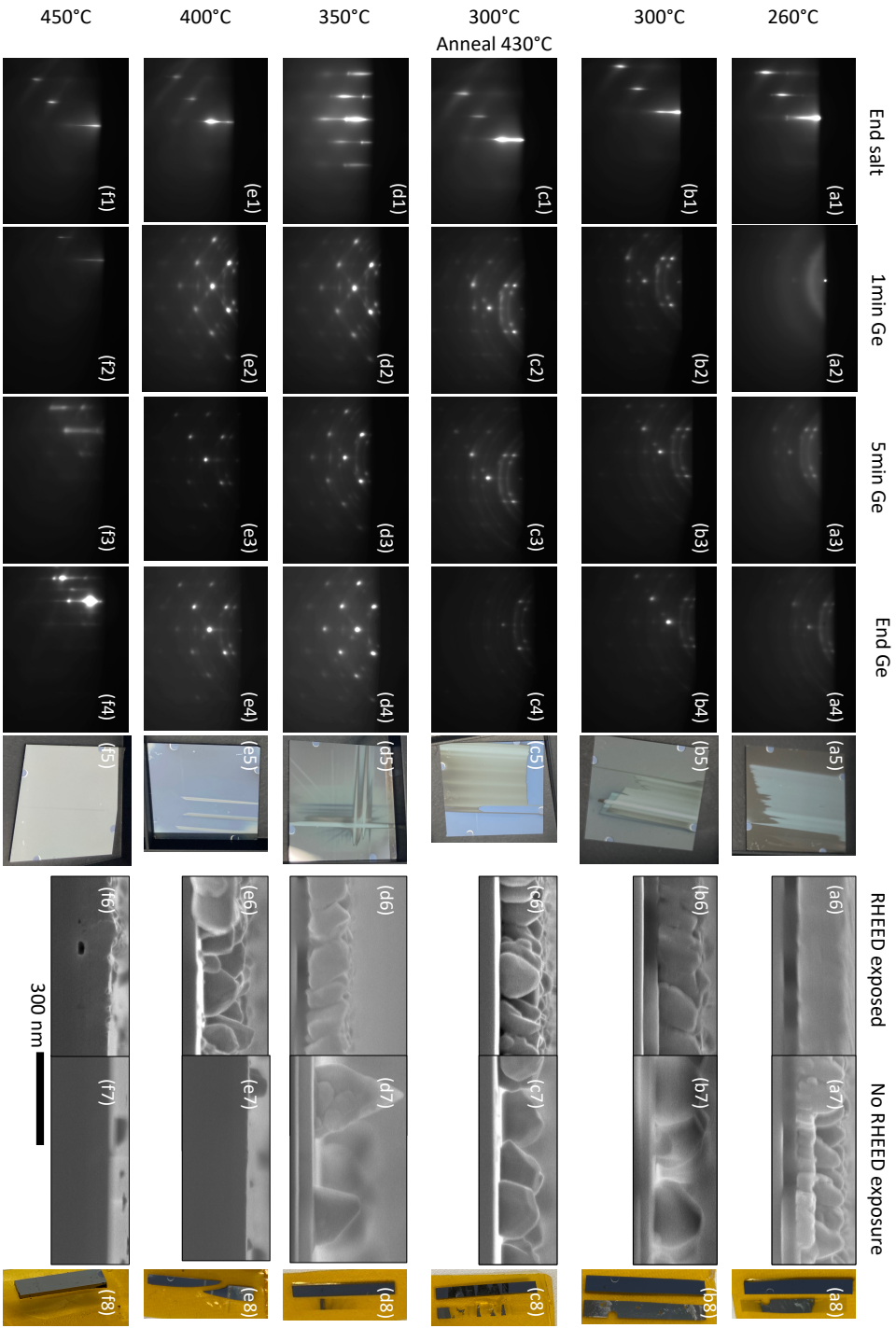

Figure S6: 20 minute Ge depositions on NaCl films on GaAs substrates with deposition beginning at 150°C and continuously until the Ge nucleation is started. (a) ~20min total salt followed by Ge deposition at 260°C, (b) ~23min total salt followed by Ge deposition at 300°C (c) ~23min total salt followed by Ge deposition at 300°C and annealed at 430°C for 5 minutes (d) ~24.5min total salt followed by Ge deposition at 350°C (e) ~49min total salt followed by Ge deposition at 400°C (f) 83 min total salt followed by Ge deposition at 450°C. RHEED images taken at (1) the end of NaCl deposition, just prior to (2) 1 minute into, (3) 5 minutes into, and (4) at the end of Ge deposition. (5) Images of the samples after removal from the chamber showing streaks from the presence of the RHEED beam throughout growth. (6) Ge films attached to Kapton tape after removal from the substrate via placement in water (lower) and the GaAs substrate (upper). Cross-section SEM images of areas (7) with RHEED exposure and (8) without RHEED exposure (scale bar is the same for all images)

The deposition of elemental Ge on NaCl thin films on GaAs substrates was also investigated because Ge has a similar lattice constant to NaCl and GaAs, but since it is only a single atomic species, there are no added complications with stoichiometry, flux-matching, or overpressures of another volatile species which could affect the growth mode. GaAs films have also been demonstrated on NaCl films previously with the use of Ge interlayers.[15,24] A similar growth scheme as outlined in Supplementary S2(b) is used. NaCl was deposited at low temperatures for a $t_{NaCl}$ and continuously for an additional $t_{ramp1}$ before $T_{GaAs1}$ was reached. Then once the temperature stabilized, Ge was deposited for 20 minutes. The sample was cooled immediately after Ge deposition. A series of growths was performed varying the $T_{GaAs1}$ of Ge from 260-450°C and analyzed using RHEED (along a [110] direction) and cross section SEM. The very clear discrepancy in areas with RHEED exposure (WRE) and areas with no RHEED exposure (NRE) in this more simple growth system was where we realized the importance of the presence of an electron beam.

For the sample with $T_{GaAs1}$=260°C (Figure S6(a)) the NaCl pattern prior to deposition is still bright and streaky. Upon opening the Ge shutter the pattern quickly dims and after 60s (Figure S6(a2)) the pattern shows diffuse rings meaning the film is highly polycrystalline. After 5 minutes, the wide blurry rings begin to give way to more discrete rings and spots (Figure S6(a3)); but with more time the pattern does not change, except for an overall slight dimming. The initial nucleation is likely highly polycrystalline, and at these low temperatures possibly somewhat amorphous as well, and although further deposition seems to suggest a more crystalline layer, the grains are very small and randomly oriented. The 5$^{th}$ column shows images of the sample after removal from vacuum, and here the RHEED exposed area can be seen clearly in the central region. Cross section SEM for areas with RE and with NRE are shown in Figures S6(a6 and a7), respectively. The area WRE shows the darker NaCl film completely covered by a dense Ge film. In the area with NRE there is still a NaCl layer of similar thickness. However, the Ge overlayer, while still mostly covering the NaCl, is very rough with significant variations in height. One thing that was not mentioned in the main text as it is outside the scope of this work is the ability of the NaCl to function as a water-soluble release layer. The final column of this figure (Figures S6(a8-f8)) shows a piece of the sample that was cleaved off and had the Ge attached to Kapton tape. The substrate and tape were submerged in water, where the NaCl would rapidly dissolve allowing for near-immediate removal of the Ge overlayer. Figure S6(a8) clearly shows the central (area WRE) lifted off. The edge areas are less dark because the thickness of the Ge is still transparent, but this region also lifted off.

Similar growth and analysis of a sample with $T_{GaAs1}$=300°C was performed (Figures S6(b)). The starting NaCl is similar to the previous sample, but the initial Ge deposition no longer shows blurry rings. Instead, the pattern is very spotted with more discrete rings. Thus, when grown at this temperature the Ge has a higher degree of short-range ordering. Further growth results in dimming of the rings and the shadow spots, while the primary spots (which would match up with NaCl) become brighter (Figure S6(b4)). This suggests that with further growth a more textured film is being achieved, and with further growth maybe even something approaching single-crystalline could be achieved. Because it would be nucleated on material consisting of multiple orientations, it would likely have a high defect density. Again, looking at the sample after removal from the chamber the area WRE can be clearly differentiated in the center (Figure S6(b5)). Cross section SEM shows that the film in the central area WRE is not quite a cohesive film, and although the thickness is similar to the previous case, there is the appearance of discrete columnar islands. In the area with NRE, the islands are much more discrete and further apart. But again, with persistent NaCl layers beneath, the tape peeled easily from the substrate either with or without placement in water and the Ge can be seen still stuck to the tape (Figure S6(b8)).

The previous sample is repeated again, except with an additional anneal at 430°C for 5 minutes in an effort to coalesce islands and achieve more crystalline material (Figures S6(c)). Unsurprisingly, the first three RHEED patterns look nearly identical as the growths are identical up until this point. However, the pattern after the annealing (Figure S6(c4)) dimmed considerably, with no apparent reduction in the relative brightness of the rings or shadow spots. Signifying that the crystallinity was not improved and the film just got rougher. The image of the sample after removal from the chamber shows the left half of the sample exposed to RHEED clearly. SEM no longer shows the presence of a NaCl layer in either area. Instead, Ge islands are simply sitting on the GaAs substrate surface. The density of islands in the area WRE is higher than the area with NRE, but because neither area had a conformal film to protect the NaCl for heating to high temperatures, the NaCl desorbs away. When attempting to remove the Ge islands from the substrate (Figure S6(c8)) it was more sporadic. Perhaps some NaCl existed some places that were not imaged in SEM, facilitating removal, while in other regions the Ge fused to the substrate and could not be lifted off.

Another sample was grown, now increasing $T_{GaAs1}$ to 350°C (Figures S6(d)). The RHEED pattern of the NaCl is again bright and streaky. But after the first minute of Ge deposition, the pattern is slightly different than previous cases. There are not full rings, rather they are broken into spots, often intersecting the symmetric shadow spots about the primary. The primary spots are very bright and show chevrons connecting them. This indicates a more textured film than the previous cases. With further growth, the primary spots get relatively brighter compared to the rings and shadow spots as well signifying an improvement of the crystal structure. The lines from the RHEED exposure can be seen again, including marks from when the sample was rotated 90° during growth (Figure S6(d5)). The SEM images still show persistent NaCl layers in both areas WRE and NRE (Figures S6(d6,7)). In the area WRE the Ge more closely resembles a film of approximately the target thickness, while outside of the RHEED-exposed area the density of the islands is much lower. These discrete islands also can be much taller than expected and are extremely faceted. This suggests that the Ge adatoms can move around on the NaCl surface until they find an existing Ge island to help grow. When attempting to remove the overlayer from the substrate as shown in Figure S6(d8), the tape peeled easily from the substrate. There is a clear line where the area was exposed to the RHEED. Moving away from that, the dark color fades. This is because where the RHEED was is the dense film, and moving away from this region the density of the Ge islands decreases, until they are no longer visible with a camera.

The RHEED patterns when increasing $T_{GaAs1}$ to 400°C (Figures S6(e1-4)) are similar to the previous case. And while streaks can be seen on the sample from the presence of the RHEED beam (Figure S6(e5)), the SEM images reveal discrete islands (Figure S6(e6)), and nothing resembling a full film. The nucleation rate of Ge on NaCl at these temperatures must be lower, and the NaCl is definitely desorbing faster. Even with a longer initial NaCl deposition ($t_{NaCl}$) of 30 min, there is no NaCl present underneath these Ge islands. Additionally, in the area with NRE, neither a NaCl layer nor Ge islands are really observed. It is likely that in this area the nucleation density of Ge is so low that the NaCl fully desorbs and heteroepitaxial Ge is deposited directly on the GaAs. Attaching the sample to tape was not able to remove any material that was visible to the eye.

A final sample with $t_{NaCl}$=60 minutes was done using a $T_{GaAs1}$=450°C (Figures S6(f)). The RHEED pattern of the NaCl layer at this temperature is still streaky but starting to blur, which is understandable because the substrate temperature is almost equivalent to the effusion cell temperature. After closing the NaCl and opening the Ge for 60 seconds, the pattern has dimmed further as the sample was roughening as Ge was going down on an actively desorbing surface. With continued growth, the pattern begins to reveal spotty streaks and eventually finishes with a bright pattern that has rotational symmetry (Figure S6(f4)). However, unlike the previous samples the area WRE is barely discernable in Figure S6(f5). Cross section SEM reveals why; in the area WRE it appears only slightly rougher than the area with NRE. Small voids are present under the surface at a depth approximately equal to the target Ge layer. This would suggest that the area with NRE (Figure S6(f7)) is heteroepitaxial Ge on GaAs, and that the layer on top of the voids in the area WRE is also Ge that formed only fast enough for some NaCl to sublime out leaving behind a pore. Attempted removal of material from this substrate (Figure S6(f8)) unsurprisingly yielded no results.

## S7. Details of low temperature As-adsorption with RHEED images

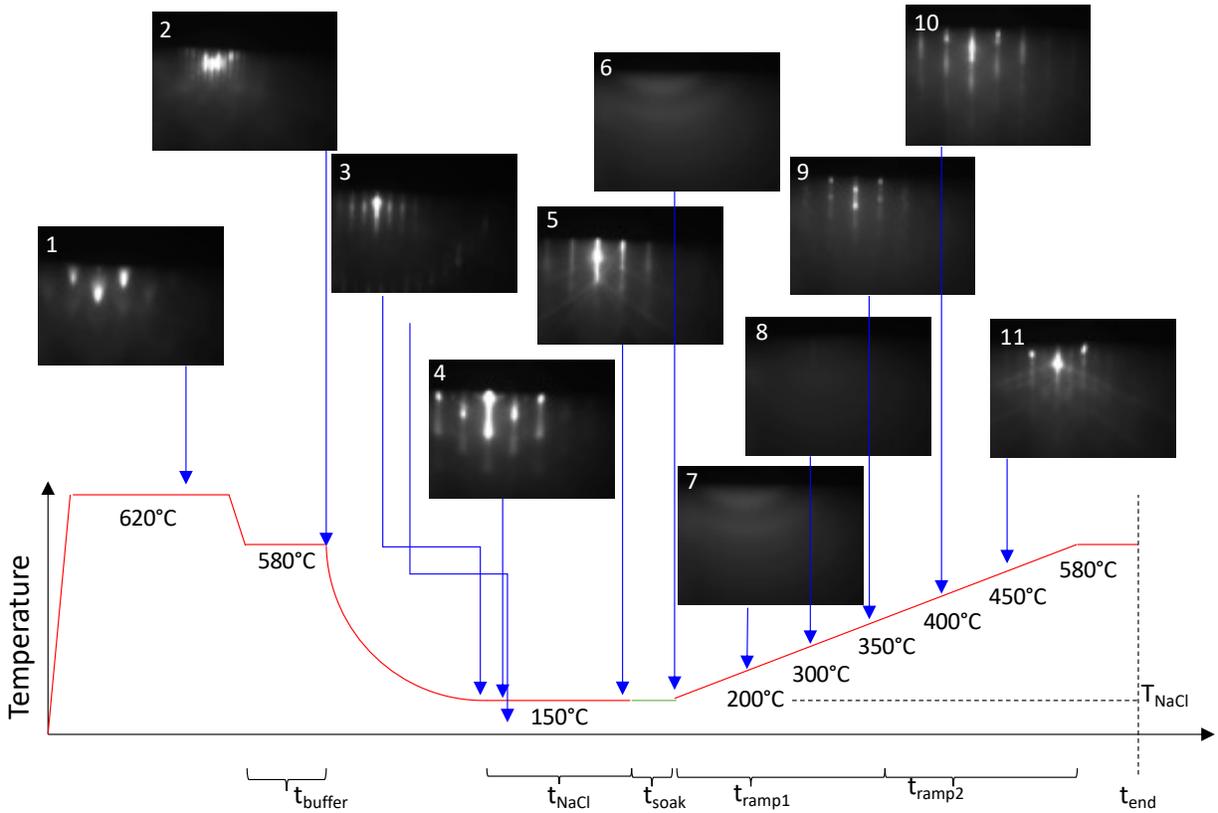

*Figure S7: Growth schematic and RHEED images at different points throughout the growth of a sample using RHEED-assisted As-adsorption*

This section pertains to the growths in section 2.3.2 in the main text and will give discussion around the electron beam induced As-adsorption. A representative growth schematic with RHEED images at different points throughout the growth labeled with numbers 1-11. The first three steps are identical for all other growths in this study, moving sequentially through the growth, we begin with a 25 minute oxide desorb and the first 3 steps are the same

1. After a few minutes at 620°C under a high As background pressure ($6.9 \times 10^{-6}$ torr), the GaAs surface which was once covered with an oxide and did not give a good RHEED pattern, now shows a bright pattern typical of an oxide free surface. The sample is cooled to 580°C and a 300 nm GaAs buffer is grown to clean up the surface.
2. The RHEED now shows a bright streaky 2×4 pattern (shown here along the $<\bar{1}10>$) at the end of the buffer layer signifying a smooth, well-ordered and As-stabilized surface. The sample is kept under a lower As background pressure ($\sim 1.0 \times 10^{-6}$ torr) while cooling until ~340°C before closing the As and cooling fully to $T_{NaCl}$.
3. A c(4×4) reconstruction appears with cooling under a constant As-flux at intermediate temperatures and persists down to $T_{NaCl}$ without losing any broadening of dimming of the peaks.

The following two images are of NaCl deposition at $T_{NaCl}$=150°C, however the trends observed are relatively similar for other $T_{NaCl}$ shown in the main text of this study. More details on the NaCl deposition can be found elsewhere.[25]

4. Upon opening the NaCl shutter the GaAs reconstruction disappears quickly, and the pattern is replaced by spotty streaks. There are also faint rings present during the early stages. This suggests that during the formation of the first few MLs of deposition the NaCl is not perfectly smooth and has some degree of polycrystallinity.
5. With further NaCl deposition the streaks become tighter and Kikuchi lines become visible. The surface becomes smoother and appears single crystalline with the spacing between streaks similar to that of the original GaAs.

The following steps pertain only to growths with the RHEED induced As-adsorption.

6. The NaCl shutter is closed and the temperature is held constant. An As flux= $1.22 \times 10^{15}$ atoms/cm$^2$/s (matching that of the Ga-flux used for 33 nm/min deposition rates) is supplied to the surface. Under constant RHEED exposure, the pattern transitions over the course of a few seconds from streaky (5) to diffuse (6) suggesting the condensation of amorphous material. However, when the location of the RHEED beam is moved pattern 5 (bare NaCl) is again observed, but it immediately begins to fade until a pattern matching (6) is observed. This stepping of the RHEED beam (which is ~1mm wide) is repeated until the amount of desired area is covered and appears diffuse. Examples of the stark differences caused in areas scanned by the RHEED can be seen in the images of 2×2 cm samples pictured in Supplementary sections 6 and 8.
7. Once the desired amount of surface is exposed to the RHEED beam, the sample is heated (at a rate of 20°C/min) under the same As-flux. The pattern remains unchanged with heating to 200°C (7).
8. Heating further to 300°C, a very faint primary streak starts to appear.
9. Once the temperature reaches 350°C (9) the pattern reveals obvious spotty streaks that match perfectly with the NaCl. Excess As desorbs from GaAs at ~320°C, thus we believe the same thing is happening with material on a NaCl surface. However, one must be careful because any additional exposure of the RHEED beam at temperatures <320°C can re-condense amorphous material, leading to the false assumption that there is still amorphous material everywhere, when in reality areas that were not re-exposed have fully desorbed all condensed material and would already display a pattern similar to (9).
10. By the time the sample is heated to 400°C the spots present at lower temperatures have transitioned into a brighter, streakier pattern. Even though NaCl adsorption is high at these temperatures, cross-section SEM images of samples where GaAs deposition begins at this temperature (Figure 8 main text) reveals persistent NaCl layers. Thus, this pattern is representative of a NaCl surface.
11. Further heating to 450°C shows the brightening of the primary spot, and the reappearance of reconstructions. Because of the reintroduction of the surface reconstructions, we believe that in this pattern is now representative of the GaAs surface. It is worth noting that the $t_{NaCl}$ for this growth is 10 minutes, and the NaCl thickness is only ~30 nm and does not take long to desorb. Cross section SEM in this case shows only smooth GaAs and no evidence that a NaCl layer was ever deposited.

   If $t_{NaCl}$ is increased to 60 minutes (~180 nm) the RHEED pattern does not yet show reconstructions at this point because not all of the NaCl has desorbed. Despite the RHEED signifying the presence of NaCl at the onset of nucleation at 450°C in this case, there is no observable NaCl layer via SEM measurements. Instead, small holes are observed at the interface. Thus, because the NaCl is still so volatile, it fully desorbs prior to coalescence of a full GaAs film to protect it.

## S8. Quantitative pole figures

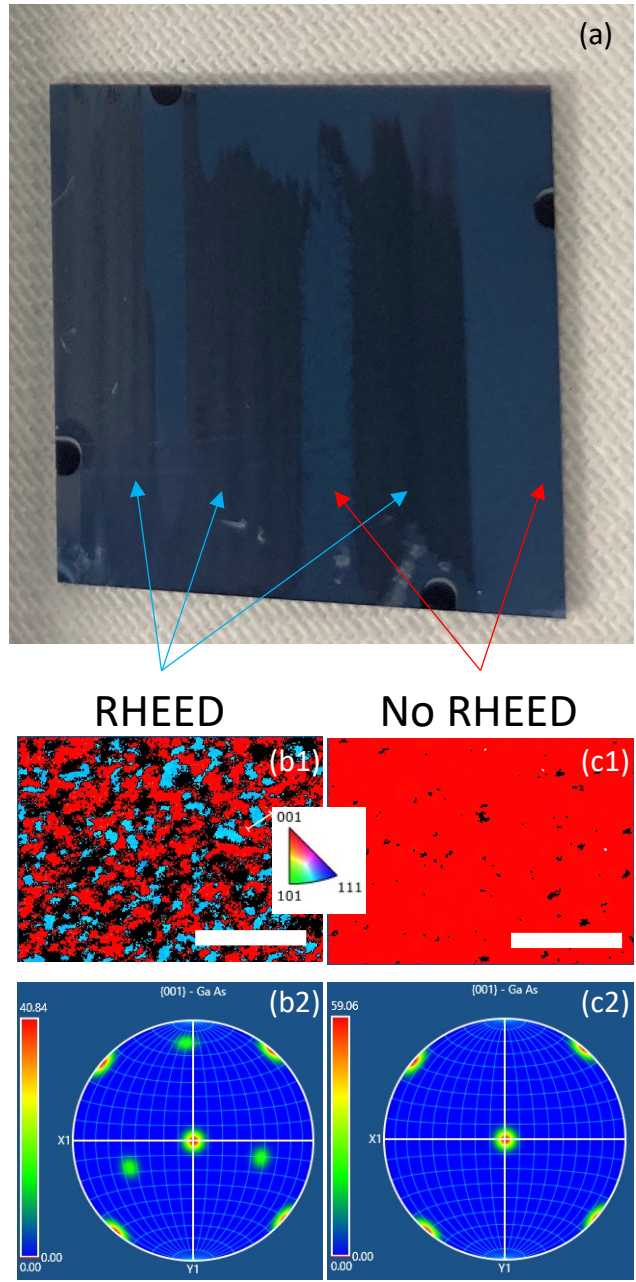

*Figure S8: (a) Image of sample from Figure 8 in the main manuscript with (1) EBSD orientation maps and (2) the corresponding pole figure for the (b) RHEED exposed areas and (c) areas without RHEED exposure.*

This section serves to highlight and support the data contained in Figure 8 in the main text. Figure S8(a) shows an image of the sample after removal from the vacuum chamber. When inside the chamber, the RHEED beam was incoming from the direction perpendicular to the top edge of the sample. The area exposed to the RHEED appears darker marked by the light blue lines. The red lines point out areas that were not exposed to RHEED, either deliberately or because the glancing angle beam was blocked by the sample holder clips or flakes on the sample holder. In the absence of any flakes present on the holder (as was observed with other samples), the beam can be scanned across the entirety of the sample during the As-adsorb step to provide a relatively uniform large area. The presence of a shadow from the sample holder clip is not avoided with a single scan, but if the sample were rotated 90° and the beam moved across again it could cover that area. This was not attempted.

Figures S8(b1,c1) are copied from the main text, but now displayed with their respective quantitative pole figures (PFs). Both PFs show the angle of {100} planes with respect to the surface normal. The central spot (at 0°) is indicative of the red areas present in the electron back scatter diffraction (EBSD) orientation maps. The additional set of 4 spots located at 90° from the center, and ~45° from the X and Y axes correspond to the other planes of a single (001) oriented crystal. The 45° misorientation is because the samples are mounted with the cleaved edge ({110}) parallel to the sample holder edge (which is used as the x-y reference). A completely red ((001) up) oriented surface could also be observed with each grain being azimuthaly misoriented (fiber oriented). In a case such as this a full ring would be observed around the outside if there was zero rotational registry with the underlying material, or multiple sets of spots could be observed around the edge if specific rotations were preferred. Here, only a single set of 4-fold rotationally symmetric spots are observed, and because the samples are 0±0.1° offcut, we can say these grains are exactly aligned with the substrate.

As discussed in the main text, the RHEED exposed area in Figure S8(b2) shows another color (light blue) corresponding to {221} grains. The PF of this region shows a single set of spots correspnding to only a single rotational domain for these grains. The origin of this orientation, why there is only a single rotation, and how to remove them to move toward more single-crystal material is the subject of separate work.

## References

[15] S. Sharma, C.A. Favela, S. Sun, and V. Selvamanickam, Conference Record of the IEEE Photovoltaic Specialists Conference **2020-June**, 0744 (2020).

[24] A.J. Shuskus and M.E. Cowher, *Fabrication of Monocrystalline GaAs Solar Cells Utilizing NaCl Sacrificial Substrates* (1984).

[25] B.J. May, J.J. Kim, P. Walker, H.R. Moutinho, W.E. McMahon, A.J. Ptak, and D.L. Young, Journal of Crystal Growth (in-review).

[26] F. Tang, T. Parker, G.C. Wang, and T.M. Lu, Journal of Physics D: Applied Physics **40**, (2007).